\documentclass[11pt]{article}
\usepackage{jheppub}
\usepackage{braket}
\usepackage{graphicx}
\usepackage{caption}
\usepackage{subcaption}

\def\tr{{\rm tr}}
\def\CA{{\cal A}}

\def\CD{{\cal D}}
\def\CE{{\cal E}}
\def\CF{{\cal F}}

\def\CH{{\cal H}}

\def\CK{{\cal K}}

\def\CR{{\cal R}}
\def\CM{{\cal M}}
\def\CN{{\cal N}}
\def\CO{{\cal O}}

\def\BH{\mathbb{H}}
\def\BR{\mathbb{R}}
\def\BS{\mathbb{S}}
\def\BT{\mathbb{T}}
\def\BZ{\mathbb{Z}}

\def\be{\begin{eqnarray}}
\def\ee{\end{eqnarray}}

\def\tr{\operatorname{tr}}

\def\Vol{\operatorname{Vol}}

\def\tr{\operatorname{tr}}

\def\d{{\rm d}}

\def\action{I}

\title{The Edge of Entanglement: Getting the Boundary Right for Non-Minimally Coupled Scalar Fields}

\author[a]{Christopher P. Herzog}
\author[b]{and Tatsuma Nishioka}

\affiliation[a]{C. N. Yang Institute for Theoretical Physics, Department of Physics and Astronomy\\
Stony Brook University, Stony Brook, NY 11794, USA}

\affiliation[b]{Department of Physics, Faculty of Science,
The University of Tokyo\\
Bunkyo-ku, Tokyo 113-0033, Japan}

\abstract{
In entanglement computations for a free scalar field with coupling to background curvature, there is a boundary term in the modular Hamiltonian which must be correctly specified in order to get sensible results.  We focus here on the entanglement in flat space across a planar interface and (in the case of conformal coupling) other geometries related to this one by Weyl rescaling of the metric.  For these ``half-space entanglement'' computations, we give a new derivation of the boundary term and revisit how it clears up a number of puzzles in the literature, including mass corrections and twist operator dimensions. 
We also discuss how related boundary terms may show up in other field theories.

}

\preprint{UT-16-30,
YITP-16-40}

\begin{document}
\maketitle

\section{Introduction}
\label{sec:intro}

Measures of quantum entanglement, and especially entanglement entropy, have been an inspirational and unifying theme in theoretical physics in recent years.  Invented largely in the quantum information community, they have been successfully applied in a much broader context.  That entanglement entropy and black hole entropy both have an area law \cite{Bombelli:1986rw,Srednicki:1993im} has been a source of inspiration in the quantum gravity community.  Certain types of entanglement entropy are believed to order quantum field theories under renormalization group flow \cite{Casini:2006es,Casini:2012ei,Solodukhin:2008dh,Komargodski:2011vj}.  These entanglement measures can also serve as order parameters for certain exotic phase transitions \cite{Osborne:2002zz,Vidal:2002rm,Kitaev:2005dm,Levin:2006zz}.

In our field theory context, we are interested in the entanglement between a spatial region $A$ and its complement $\bar A$.  We assume, for now, that the Hilbert space may be factorized with respect to these spatial regions: $\CH = \CH_A \otimes \CH_{\bar A}$.  Given this nontrivial assumption, one may construct the reduced density matirx $\rho_A = \tr_{\bar A} \rho$ by tracing over the degrees of freedom in the complementary region $\bar A$, where $\rho$ is the initial density matrix.  The R\'enyi entropies are then moments of the reduced density matrix,
\be
S_\alpha \equiv \frac{1}{1-\alpha} \log \tr [(\rho_A)^\alpha] \ ,
\ee
while the entanglement entropy itself can alternately be defined as the $\lim_{\alpha \to 1} S_\alpha$ of the analytically continued R\'enyi entropies or as the
von Neumann entropy with respect to the reduced density matrix:
\be
S_E \equiv - \tr (\rho_A \log \rho_A) \ .
\ee
The modular Hamiltonian is defined to be the logarithm of the reduced density matrix,
\be
H \equiv -\log \rho_A \ ,
\ee
and plays a key role in computing entanglement and R\'enyi entropies as well as other measures of quantum entanglement such as the mutual information and the relative entropy.

In general, $H$ is non-local.  One simple and important case where $H$ is believed to be known is in computations of the entanglement across a planar interface $x^1=0$ in flat space coordinatized by $x^\mu$, $\mu = 0, 1, \ldots, d-1$.  In this case, by the Bisognano-Wichmann theorem \cite{Bisognano:1976za,Bisognano:1975ih}, the modular Hamiltonian is given by the generator of boosts in the direction $x^1$ perpendicular to the interface.  In the Euclidean setting, where $x^1$ and the Euclidean time coordinate $i x^0$ form a plane, these boosts are simply rotations about the origin.  In terms of the stress tensor $T_{\mu\nu}$ (defined by varying the action with respect to an external metric), we then naively should be able to claim that the modular Hamiltonian is
\begin{align}\label{Hdef}
H_{\rm cov} = 2 \pi \int  \d^{d-1} x \, x^1 \, T_{00}(x)   \ .
\end{align}
We use the subscript ``cov'' for covariant.
For conformal field theories, this result has additional implications for geometries related to a planar interface in flat space by Weyl rescaling of the metric.  For instance, the modular Hamiltonian for entanglement across a spherical interface in flat space would then also be known (as described for example in refs.\ \cite{Casini:2011kv,Wong:2013gua}).

The problem with the definition (\ref{Hdef}) is that the charge $H$ can be modified by a boundary term \cite{Barnich:2001jy,Fursaev:1998hr,Frolov:1997up}.
More generally, given a conserved current $J^\mu$, such that $\partial_\mu J^\mu = 0$, we are free to introduce an antisymmetric tensor field $q^{\mu\nu} = - q^{\nu\mu}$ and add a derivative of it to the current $J^\mu \to J^\mu + \partial_\nu q^{\nu\mu}$ without spoiling the conservation condition.  The conserved charge is then modified by the boundary integral $\int \d^{d-2} x \, q^{01}$.   (With regards to $H$, $q^{\mu\nu}$ could for example be a time component of the Belinfante tensor.)

In the case of the conformally coupled scalar field $\phi$, 
refs.\ \cite{Herzog:2014fra,Lee:2014zaa} proposed seemingly distinct methods for fixing the ambiguity.  Ref.\ \cite{Herzog:2014fra}'s proposal, which we henceforth adopt in this paper, is to add a boundary term such that  
\be
\label{Hfixed}
H = 2 \pi \int_{x^1 > \epsilon}  \d^{d-1} x\, x^1 \, T_{00}(x)    + 2 \pi \xi \int_{x^1 = \epsilon} \d^{d-2} x \, \phi^2 \ ,
\ee
where $T_{\mu\nu}$ is the usual ``improved'' stress tensor for a conformally coupled scalar field, derived by varying the action with respect to the metric, and $\xi = \frac{d-2}{4 (d-1)}$ is the conformal coupling.
A nice feature of this expression is that each term is separately Weyl invariant; it is straightforward to construct $H$ for geometries related by Weyl rescaling.  In contrast, ref.\ \cite{Lee:2014zaa} proposed that one should use the ``unimproved'' stress tensor, dropping the $\nabla^2 \phi^2$ term from the $T_{00}$ component of the ``improved'' stress tensor.  In this case, it is less clear how to apply a Weyl rescaling.  Ref.\ \cite{Casini:2014yca} later partially clarified the situation by pointing out that at least in flat space, the $\nabla^2 \phi^2$ contribution could be integrated by parts to yield the boundary term in (\ref{Hfixed}).  
In section \ref{sec:bryterm}, we present two partially independent arguments for the choice of boundary term in (\ref{Hfixed}).\footnote{%
 In related work discussing Wilson-Fisher fixed points and the $O(N)$ model, ref.\ \cite{Metlitski:2009iyg} found evidence that such a boundary term could be dynamically generated through renormalization group flow.  This paper focuses on the Gaussian fixed point, and we have little to say about the effect of interactions on entanglement entropy.
}

As suggested by refs.\ \cite{Herzog:2014fra,Lee:2014zaa} and as we further clarify here, the $\phi^2$ boundary term resolves a number of puzzles in the literature concerning various perturbative corrections to entanglement entropy of a conformally coupled scalar.  
The unifying feature of these puzzles is that the perturbative result depends on correlation functions involving $H$.  If the boundary term in (\ref{Hfixed}) is not included, these correlation functions take the wrong value, and the perturbative result will disagree with other independent computational methods.  
Ref.\ \cite{Herzog:2014fra} discovered this boundary term affected  thermal corrections to entanglement entropy.  Ref.\ \cite{Lee:2014zaa} noted the importance of this term for computing the entanglement entropy of a massive scalar perturbatively in the mass.  
Ref.\ \cite{Lee:2014zaa} also noted its importance for computing R\'enyi entropies in the $\alpha \to 1$ limit, or equivalently the dimension of twist operators perturbatively in an $\alpha \to 1$ limit.  
(The discrepancy in the $\alpha \to 1$ limit of twist operators was noted but not explained in ref.\ \cite{Hung:2014npa}.)  
In sections \ref{sec:mass} and \ref{sec:twist}, we extend to all dimension $d \geq 3$ the perturbative mass and twist operator calculations presented in ref.\ \cite{Lee:2014zaa} for the case of a dimension $d=3$ scalar.  (For $d=2$, the conformal coupling vanishes, $\xi = 0$, and there is no discrepancy to be explained.)

Indeed we claim for a massive scalar with an arbitrary non-minimal coupling, $\xi \neq \frac{d-2}{4(d-1)}$, the improved modular Hamiltonian (\ref{Hfixed}) continues to be the correct one to use in ``half-space'' entanglement calculations.  With a mass, we lose the ability to perform Weyl transformations, but we may still consider the mass dependence of the entanglement across the $x^1=0$ interface.    
The operator algebra for a non-minimally coupled scalar field in flat space cannot depend on $\xi$ because the curvature vanishes.  Therefore, it seems reasonable to assume that the entanglement entropy cannot depend on $\xi$ \cite{Casini:2014yca}.  For instance in flat space, standard numerical methods which depend on the operator algebra \cite{Srednicki:1993im} cannot detect a nonzero value of $\xi$.   We show that the boundary term in (\ref{Hfixed}) acts precisely to cancel any naive $\xi$ dependence in the area law term of the entanglement entropy for a massive scalar.   In this context, the calculations we present are very similar to those in ref.\ \cite{Casini:2014yca}, but our interpretation is somewhat different.  
The possibility and attendant difficulties of $\xi$-dependence in entanglement entropy is a twenty year old subject \cite{Larsen:1995ss, Solodukhin:1995ak,Solodukhin:1996jt,Hotta:1996cq} which has still not been successfully resolved \cite{Akers:2015bgh}.  We hope our arguments may finally help put the debate to rest.  

The paper is organized as follows.  In section \ref{sec:bryterm}, we give two independent arguments for the necessity of the boundary term in the modular Hamiltonian (\ref{Hfixed}).  Both arguments rely on specifying appropriate boundary conditions for the entangling surface.  The first argument uses the modular Hamiltonian, while the second uses the path integral on a conical space.  
The next three sections discuss mass contributions to the entanglement entropy of a non-minimally coupled scalar; we discuss in order the small mass limit, the large mass limit, and the case of arbitrary mass.
In section \ref{sec:mass}, we explain how the boundary term resolves a discrepancy in perturbative calculations of the mass dependence of the entanglement entropy for a conformally coupled scalar.  
As we are in the business of clearing up puzzles in the calculation of entanglement entropy for non-minimally coupled scalars, we decided to resolve a puzzle \cite{Banerjee:2015tia} that turns out not to be associated with boundary terms.  
This puzzle concerns the large mass limit in 2+1 dimensions.  As we describe in section \ref{sec:largemass}, the resolution of this puzzle has a more mundane origin in the dimensional dependence of the conformal coupling.  Third, we discuss the mass dependence of the area law contribution for an arbitrary mass, non-minimally coupled scalar in section \ref{ss:universal}, arguing that the prescription (\ref{Hfixed}) should hold also away from the conformal coupling $\xi = \frac{d-2}{4(d-1)}$.
Moving on, in section \ref{sec:twist}, we use the boundary term to clear up a discrepancy in the calculation of twist operator dimensions (or equivalently R\'enyi entropy in the limit $\alpha \to 1$).
The Discussion in section \ref{sec:discussion} is an attempt to apply the lessons learned here about boundary terms to gauge fields and more general quantum field theories.  
Appendix \ref{sec:numerics} contains details of the numerical algorithms we use to check our results.  Appendix \ref{sec:further} has two additional versions of the perturbative mass calculations described in section \ref{sec:mass}.

\section{Fixing the Boundary Term Ambiguity}
\label{sec:bryterm}

The action for a non-minimally coupled, massive scalar field $\phi$ in a $d$-dimensional curved space time $\CM$ with boundary $\partial \CM$ is
\begin{align}
\label{ScalarAction}
\action = - \frac{1}{2} \int_\CM \d^d x \sqrt{-g}\, [ (\partial \phi)^2 + m^2 \phi^2 + \xi\, \CR\, \phi^2 ] - \xi \int_{\partial \CM} \d^{d-1} x \sqrt{-\gamma}\, \CK \,\phi^2 \ .
\end{align}
The action is Weyl invariant for the choices $m^2 =0$ and 
\be
\xi = \xi_c \equiv \frac{d-2}{4 (d-1)} \ .
\ee
The quantities $\CR$ and $\CK$ are the Ricci scalar curvature and the trace of the extrinsic curvature respectively.  We have also introduced the induced metric $\gamma^{\mu\nu}$ on the boundary $\partial \CM$.
The usual action has been supplemented by a boundary term $\CK\, \phi^2$ that preserves the Weyl invariance in the presence of $\partial \CM$.  
The boundary term also improves the behavior of the action under variations with respect to the metric,
which yields the usual ``improved'' stress tensor with a less well-known boundary contribution:
\begin{align}\label{ScalarST}
T^{\mu\nu} &\equiv  \frac{2}{\sqrt{-g}} \frac{\delta I}{\delta g_{\mu\nu}} \ ,\nonumber \\
&= (\partial^\mu \phi) (\partial^\nu \phi)  - \frac{1}{2} g^{\mu\nu} \left[(\partial \phi)^2 + m^2 \right]+ \xi \left(\CR^{\mu\nu} - \frac{1}{2}  \CR\, g^{\mu\nu}  -\nabla^\mu \nabla^\nu + g^{\mu\nu} \nabla^2 \right)  \phi^2 \nonumber \\
& \qquad + \xi \delta(x_\perp)\left[ (\CK^{\mu\nu} - \CK\, \gamma^{\mu\nu} ) \phi^2 - (\partial_\sigma \phi^2) n^\sigma \gamma^{\mu\nu} \right]  \ .
\end{align}
Here $n^\mu$ is an outward pointing unit normal vector to $\partial \CM$ and $\CK^{\mu\nu}$ is the extrinsic curvature.

\subsection{Hamiltonian approach}
We begin our entanglement entropy computations with the case where $\CM$ is the Rindler wedge of Minkowski space, $(x^1)^2 - (x^0)^2 > \epsilon^2$.  The cut-off $\epsilon >0$ is introduced to control singular behavior in the entanglement entropy in the limit $\epsilon \to 0$.
We would like to compute the entanglement entropy of the half space $x^1 > \epsilon$ at $x^0=0$.  This quantity is controlled by the modular Hamiltonian $H$, also known as the boost generator in the $01$-plane.
Starting from the stress tensor and the naive definition (\ref{Hdef}) of the modular Hamiltonian, $H$ should be
\be
\label{Hcov}
H_{\rm cov} &=& \pi \int_{x^1>\epsilon} \d^{d-1} x  \, x^1 \left[ \Pi^2 + (\partial_i \phi)^2  + m^2 \phi^2 \right] - 2 \pi \xi \int_{x^1 >\epsilon} \d^{d-1} x \, x^1 \vec \nabla^2 \phi^2 \ ,\\
&=&\pi \int_{x^1>\epsilon} \d^{d-1} x  \, x^1 \left[ \Pi^2 + (\partial_i \phi)^2  + m^2 \phi^2 \right] - 2 \pi \xi \int_{x^1 =\epsilon} \d^{d-2} x \,  \phi^2 \ ,
\ee
where the canonical momentum $\Pi = \partial_0 \phi$ and $\vec \nabla^2 = \Box^2 + \partial_0^2$.
In deriving the first line, we have used that $\CK_{\mu\nu} - \CK \gamma^{\mu\nu}$ vanishes, because $\CK_{\mu\nu}$ has only one nonzero component on the boundary.  We have also assumed that $\phi$ is not divergent at $x^1 = 0$, allowing us to drop a $x^1 \partial_{x^1} \phi^2$ term at the boundary.
This modular Hamiltonian evolves the theory not in $x^0$ but in $\tau =  \tanh^{-1} (x^0/x^1)$.  We have multiplied $H_{\rm cov}$ by a factor of $ 2\pi$ associated with the periodicity of the Euclidean time $\tau_E = i \tau$ or correspondingly the inverse temperature.  In moving from the first line to the second line, we have integrated by parts twice and again assumed that 
 $\phi$ is not divergent at $x^1 = 0$.

The ambiguity in $H$ can be seen by instead starting in flat space and deriving the modular Hamiltonian using Noether's theorem and boost symmetry.
The Noether charge is basically the canonical Hamiltonian multiplied by an extra factor of $x^1$:
\be
\label{Hcan}
H_{\rm can} &=& \pi \int_{x^1>\epsilon} \d^{d-1} x  \, x^1 \left[ \Pi^2 + (\partial_i \phi)^2 + m^2 \phi^2  \right] +2 \pi \xi \int_{x^1 =\epsilon} \d^{d-2} x \,  \phi^2 \ ,
\ee
where we need to be careful to evaluate $\epsilon\, \CK = 1$ before taking the limit $\epsilon \to 0$.  The subscript ``can'' here means canonical.
Clearly the expressions (\ref{Hcov}) and (\ref{Hcan}) differ by a boundary term.

The claim of refs.\ \cite{Herzog:2014fra,Lee:2014zaa,Casini:2014yca} can be paraphrased thus -- the modular Hamiltonian relevant for an entanglement entropy computation is the one with no boundary term at all,
\be\label{Improved_H}
H &=& \pi \int_{x^1>\epsilon} \d^{d-1} x  \, x^1 \left[ \Pi^2 + (\partial_i \phi)^2  + m^2 \phi^2 \right]  \ ,
\label{Hscalarcon}
\ee
or equivalently eq.\ (\ref{Hfixed}).
 
 One way of justifying this choice is through a consideration of boundary conditions.
Hamilton's equations reduce to
\be
\dot \phi &=& z \Pi \ , \\
\dot \Pi &=& \partial_z (z \partial_z \phi) + z \partial_i^2 \phi  + z\, m^2 \phi^2 \ ,
\ee
where for ease of notation we have set $x^1 = z$.  The dot means a derivative with respect to the time associated with $H$, i.e.\ the modular time $\tau$, not $x^0$.
In deriving the $\dot \Pi$ equation of motion, we have assumed that a boundary term vanishes,
\be
\left. (z \partial_z \phi + c \phi ) \right|_{z = \epsilon} = 0 \ ,
\ee
where we have allowed for a constant $c$ depending on the strength of the $\phi^2$ boundary term.
Our assumption that $\phi$ is finite at the entangling surface sets the $z \partial_z \phi$ combination to zero.  If $c \neq 0$, then the value of $\phi$ at the entangling surface is fixed.  
 Such a Dirichlet condition is not compatible with the notion of an entanglement entropy computation where the value of the field $\phi$ should be free to fluctuate at the entangling surface.  
The remaining option is to set $c=0$ and choose the modular Hamiltonian (\ref{Hscalarcon}).
A similar argument was presented in ref.\ \cite{Herzog:2014fra} but using a Lagrangian framework and the Weyl rescaled geometries: $\BH^{d-1} \times \BS^1$ where $\BH^d$ is $d$-dimensional hyperbolic space and also $\BS^{d-1} \times {\mathbb R}$.

We would like to elevate these boundary condition considerations to a more general principle:
To compute entanglement entropy, 
we must choose a modular Hamiltonian which has sensible boundary conditions for the quantum fields at the entangling surface.  For the case of the scalar, that meant the fields should be free to fluctuate but should also not be divergent at $z=0$.  
If a naively computed modular Hamiltonian, for example $H_{\rm cov}$ or $H_{\rm can}$ above, 
does not have sensible boundary conditions, one may be able to improve it by adding boundary terms.

\subsection{Path integral approach}
\label{sec:pathintegral}

We now present an alternate derivation of the same boundary term, using instead a path integral (or replica trick) approach.  
The R\'enyi entropy $S_\alpha$ across a planar interface is related to the Euclidean partition function on the conical space $C_\alpha \times {\mathbb R}^{d-2}$ where $C_\alpha$ is an $\alpha$-sheeted cover of ${\mathbb R}^2$, branched over the origin:
\be
S_\alpha = \frac{1}{1-\alpha} \log\left[ \frac{Z_\alpha}{(Z_1)^\alpha} \right]\ .
\ee
The codimension-two conical singularity allows one to consider adding codimension-two boundary terms to the action for a conformally coupled scalar, in particular
\be
\delta \action = c \int_{x^1 = x^0 = 0} \d^{d-2} x_\perp\, \phi^2 \ ,
\ee
where $c$ is an as yet undetermined constant and $\vec x_\perp$ are the transverse coordinates on $\BR^{d-2}$ parametrizing the entangling surface.  Note that such a term does not spoil the Weyl invariance of the theory.  

On $C_\alpha$, the Ricci scalar has a delta function distribution at the conical singularity $\CR = 4 \pi (1-\alpha) \delta^2(x)$.  If we put a metric $\d s^2_{C_\alpha} =\d r^2 + r^2 \d \theta^2$ on $C_\alpha$ where $0 \leq \theta < 2 \pi \alpha$, then the distribution can be written as $\CR = -2 (1-1/\alpha) \delta (r) / r$.  
The differential operator in the equation of motion is then 
\be
\Box -  \xi\, \CR = \frac{1}{r} \partial_r r \partial_r - 2 \xi \frac{1-1/\alpha}{r} \delta(r)  + \frac{1}{r^2} \partial_\theta^2 + \partial_{\vec x_\perp}^2  + m^2\ .
\ee
(There is in principle also a distributional contribution to $\Box$, proportional to $\delta (r) r \partial_r$.  However, with our choice of boundary condition, this contribution will vanish.)
To carry out the path integral, we look for eigenfunctions of this operator using a separation of variables ansatz, $\phi(t, \vec x_\perp, r) \sim e^{i n \theta / \alpha + i k_i x^i_\perp} \varphi(r)$.  We find 
\be
\varphi'' + \frac{\varphi'}{r} - \frac{2 \xi (1-1/\alpha)}{r} \delta(r) \varphi - \frac{n^2}{\alpha^2 r^2} \varphi  = -\lambda^2 \varphi \ .
\ee

This ordinary differential equation has been well studied both in the context of cosmic strings and conformal quantum mechanics (see for example  ref.\ \cite{Kay:1990cr}).   
The general solution is
\be
\varphi = c_1 J_{|n/\alpha|} (r \lambda) + c_2 Y_{|n/\alpha|} (r \lambda) \ .
\ee
Normalizability implies that if $c_2 \neq 0$, then $n < \alpha$.  An additional constraint comes from integrating the differential equation near the origin.  For the $n>0$ case, the simplest solution is to set $c_2=0$ in which case $\varphi(0) = 0$ and the effect of the delta function vanishes.  For the $n=0$ case, the delta function produces the constraint 
\be
\lim_{r \to 0} r \varphi'(r) = 2 \xi(1-1/\alpha) \varphi(0) \ .
\ee
We can try to use this constraint to find a relation between $c_1$ and $c_2$.  
One finds however that the relation can only be solved at a small cut-off scale $r_{\star}$ and not in the strict limit $r \to 0$:
\be
\frac{c_1}{c_2} = \frac{2}{\pi} \left( \frac{1}{C} - \gamma  - \log (r_\star \lambda/2) \right) \ ,
\ee
where $C = 2 \xi(1 - 1/\alpha)$.
Using this procedure, one will find also a bound state (or tachyon) $K_0(i r \lambda)$ ($\lambda$ is pure imaginary) with energy proportional to $1/r_\star$ when $\alpha < 1$:
\be
i \lambda r_\star = 2 \exp \left( -\gamma+ \frac{1}{C} \right) \ .
\ee

In the conformal case $m=0$ and $\xi = \xi_c$, the introduction of such a scale breaks conformal invariance.  The bound state also implies the theory is not stable.  
 One could imagine truncating the $n=0$ modes from the spectrum.  But that will have the awkward consequence of forcing the field to vanish at the origin, a sort of Dirichlet condition on an entanglement problem where the origin is not a special point and the field should be free to fluctuate there.
 We propose a different resolution.  We use the freedom to add a codimension-two boundary term to the action to remove the delta function distribution from the Laplacian and restore scale invariance to the solutions:
\be
\delta \action_E = -2 \pi (1-\alpha ) \xi \int_{r=0} \d^{d-2} x \, \phi^2 \ .
\label{Ibryterm}
\ee
The sign is a bit tricky. Eq.~\eqref{Ibryterm} correctly cancels the contribution from the tip to the Ricci scalar in the Euclidean signature we use in this section.  In Lorentzian signature, $\delta I$ would have the opposite sign.

It seems most natural to define theories with arbitrary values of $m$ and $\xi$ as a smooth deformation of the Weyl invariant case.  Thus, we should keep the codimension-two boundary term (\ref{Ibryterm}), with coefficient $\xi$, that cancels the delta function contribution from the Ricci scalar.
On the cone then, our action reduces to
\be
\label{actionE}
\action_E = \frac{1}{2} \int \d^d x\,  \left[ (\partial \phi)^2 + m^2 \phi^2 \right] \ ,
\ee
and, using boosts and Noether's theorem, the modular Hamiltonian to (\ref{Hscalarcon}).  (Note that the space $C_\alpha \times {\mathbb R}^{d-2}$ has no codimension-one boundary.)

The procedure we are advocating here is equivalent to the so-called Friedrichs extension \cite{Kay:1990cr} of the Laplacian on the cone.  Such an extension is typically used in heat kernel calculations on the cone.  See for example ref.\ \cite{Kabat:1995eq}.  We are arguing that the Friedrichs extension implicitly involves adding a codimension-two boundary term to the action for a non-minimally coupled scalar. 

\section{Conformal Perturbation Theory and the Massive Scalar}
\label{sec:mass}

In this section, we consider how the mass of a scalar field affects the entanglement entropy in the limit where the mass is kept small.  We will use conformal perturbation theory.

Let $\CO(x)$ be a relevant operator of dimension $\Delta (\le d)$ in $d$-dimensional QFT.
The relevant perturbation by the operator 
\begin{align}\label{Rel_Pert}
	\action_E = \action_\text{UV} + g_\CO\, \int \d^d x \sqrt{g}\, \CO(x) \ ,
\end{align}
induces an RG flow from the UV fixed point described by the CFT with the action $I_\text{UV}$ to an IR fixed point.
Perturbative studies around the UV fixed point $(g_\CO=0)$ show entanglement entropy has an expansion in the relevant coupling $g_\CO$ of the form \cite{Wong:2013gua,Blanco:2013joa,Rosenhaus:2014woa} (see also \cite{Bhattacharya:2012mi,Nozaki:2013wia,Nozaki:2013vta} for related works)
\begin{align}\label{PertS}
	S(\lambda) = S(0) + s_1\, g_\CO  + s_2\,g_\CO^2  + O\left(g_\CO^3\right) \ ,
\end{align}
where the first order term is given by the integrated two-point function of the modular Hamiltonian $H$ and the relevant operator
\begin{align}\label{1stS}
	s_1 = -\int \d^d x\sqrt{g}\, \langle H\, \CO(x)\rangle \ .
\end{align}
If we consider half-space entanglement and were to adopt the naive expression (\ref{Hdef}) for the modular Hamiltonian,
the first order term should vanish, $s_1=0$. It would vanish because the two-point function of the stress tensor and a scalar operator is zero in CFT, as noted in this context in ref.\ \cite{Rosenhaus:2014woa}.  However, as we have already discussed, the modular Hamiltonian is not simply an integral over the stress tensor.  It also involves a boundary term (\ref{Hfixed}).  This boundary term can generically give a nonzero result for $s_1$.

If the conformal perturbation theory calculation described above is to give the correct answer, it should agree with an equivalent replica trick approach to the computation.  
In order to reduce the complexity of the problem, we place the CFT on a compact space to remove the IR divergence.
Among many choices of the IR regulator, we employ a cylinder type manifold $\BR \times \BS^{d-1}$ whose Euclidean metric is
\begin{align}\label{Cylinder_metric}
	\d s^2_{\BR \times \BS^{d-1}} = \d t^2 + R^2 (\d \theta^2 + \sin^2\theta\, \d\Omega_{d-2}^2) \ ,
\end{align}
and take the entangling surface $\Sigma$ dividing $\BS^{d-1}$ into two subregions at $t=0$ and $\theta = \theta_0$ as in \cite{Banerjee:2015tia}.
Using the replica trick, the entanglement entropy can be extracted from the partition function $Z[\CM_\alpha]$ on the $\alpha$-fold cover $\CM_\alpha$ of the manifold \eqref{Cylinder_metric} branched over the codimension-two hypersurface $\Sigma$.  In particular, we find
\begin{align}\label{EntFree}
	S = \lim_{\alpha\to 1} \partial_\alpha (F_\alpha - \alpha\, F_1) \ ,
\end{align}
where $F_\alpha \equiv - \log Z[\CM_\alpha]$ is the free energy.

Instead of working on the intricate geometry $\CM_\alpha$, it is easier to make use of a Weyl rescaling to a simpler space $\BS^1 \times \BH^{d-1}$ with metric
\begin{align}\label{S1H2}
	\d s^2_{\BS^1 \times \BH^{d-1}} = R^2 \left[ \d \tau^2 + \d u^2 + \sinh^2 u\, \d \Omega_{d-2}^2 \right] \ ,
\end{align}
whose relation to the original one \eqref{Cylinder_metric} is given by $\d s^2_{\BR \times \BS^{d-1}} = e^{-2\sigma} \d s^2_{\BS^1 \times \BH^{d-1}}$ with the conformal factor \cite{Casini:2011kv}\footnote{The map follows from the coordinate transformation
\begin{align}
		\tanh (t/R) = \frac{\sin\theta_0\sin\tau}{\cosh u + \cos\theta_0 \cos\tau} \ , \qquad \tan \theta = \frac{\sin\theta_0\sinh u}{\cos \tau + \cos\theta_0 \cosh u} \ .
\end{align}
}
\begin{align}\label{ConfFactor}
	e^{-2\sigma} = \frac{\sin^2 \theta_0}{(\cos \tau + \cos\theta_0\, \cosh u)^2 + \sin^2 \theta_0 \, \sinh^2 u} \ .
\end{align}
Accordingly, the $\alpha$-fold cover $\CM_\alpha$ is mapped to the space \eqref{S1H2} with the $\alpha$ times larger period $\tau \sim \tau + 2\pi \alpha$.
The perturbative expansion of the free energy for the relevant perturbation \eqref{Rel_Pert} is easily seen to take the form of
\begin{align}\label{FreeN}
	F_\alpha(\lambda) = F_\alpha(0) + \sum_{l=1}^\infty \frac{g_\CO^l }{l!}\int \cdots \int_{\BS^1_\alpha \times \BH^{d-1}} e^{(\Delta -d)\sigma(x_1)}\cdots e^{(\Delta -d)\sigma(x_l)} \langle \CO(x_1) \cdots \CO(x_l)\rangle_{\BS^1_\alpha \times \BH^{d-1}} \ .
\end{align}
Here $\BS^1_\alpha$ stands for a circle parametrized by $\tau$ of the period $2\pi \alpha$.
Formal substitution of \eqref{FreeN} into the formula \eqref{EntFree} will calculate the entanglement entropy, the first order term \eqref{1stS} being\footnote{%
To recast the term in this form, we utilized the replica $\BZ_\alpha$ symmetry that manifests itself in the domain of integration as $\int_{\BS^1_\alpha \times \BH^{d-1}} \cdots = \alpha \int_{\BS^1 \times \BH^{d-1}}\cdots $.
}
\begin{align}\label{OnePoint}
	s_1 = \int_{\BS^1 \times \BH^{d-1}} e^{(\Delta -d)\sigma(x)} \lim_{\alpha\to 1} \partial_\alpha  \left[ \langle \CO(x) \rangle_{\BS^1_\alpha \times \BH^{d-1}} - \langle \CO(x) \rangle_{\BS^1 \times \BH^{d-1}} \right] \ .
\end{align}
In this form, there is no particular reason for $s_1$ to vanish.  

For a concrete example, we consider a free scalar theory \eqref{ScalarAction}
with mass term $m^2\phi^2/2$.
The theory becomes conformal when $\xi = \xi_c$ and $m=0$ and we regard the mass term as a relevant perturbation with $g_\CO = m^2$ and $\CO = \phi^2/2$ of $\Delta = d-2$.
Indeed, we will see below that for a mass perturbation, $s_1$ does not vanish and that the replica trick agrees with the modular Hamiltonian approach, provided we include the $\phi^2$ boundary term in $H$.

\subsection{Free energy method}

We use the replica trick to compute the $O(m^2)$ correction to the entanglement entropy of a free massive scalar field in $d \geq 3$. 
We assume that the one-point function $\langle \CO(x)\rangle_{\BS^1_\alpha\times \BH^{d-1}}$ does not depend on the position $x$ as the space has  translational invariance $\BS^1_\alpha$ and the homogeneity on $\BH^{d-1}$.
Then one can factorize \eqref{OnePoint} to
\begin{align}
	s_1 = V_d \,\lim_{\alpha\to 1} \partial_\alpha\left[ \langle \CO(x) \rangle_{\BS^1_\alpha \times \BH^{d-1}} - \langle \CO(x) \rangle_{\BS^1 \times \BH^{d-1}} \right] \ ,
\end{align}
where $V_d$ is the integral of the conformal factor
\begin{align}
	 V_d = \int_{\BS^1 \times \BH^{d-1}} \d^d x \sqrt{g}\,e^{(\Delta -d)\sigma(x)} \ .
\end{align}
We need the expectation value of the one-point function to evaluate $s_1$, which can be read off from the partition function of the theory perturbed by the relevant operator defined on $\BS^1_\alpha \times \BH^{d-1}$.
Namely, the one-point function is obtained by taking a derivative of the free energy $F_\alpha \equiv - \log Z[\BS^1_\alpha \times \BH^{d-1}]$:
\begin{align}
	\langle \CO(x) \rangle_{\BS^1_\alpha \times \BH^{d-1}} - \langle \CO(x) \rangle_{\BS^1 \times \BH^{d-1}} = 
	\Vol\left(\BS^1_\alpha \times \BH^{d-1}\right)^{-1} \partial_{g_\CO} (F_\alpha - \alpha\, F_1)\big|_{g_\CO = 0} \ ,
\end{align}
where we again used the homogeneity assumption.
Putting all together, we end up with 
\begin{align}
\label{s1init}
	s_1 = \frac{V_d}{\Vol\left(\BS^1\times \BH^{d-1}\right)} \,\lim_{\alpha\to 1} \partial_\alpha \partial_{g_\CO} (F_\alpha - \alpha F_1)\big|_{{g_\CO} = 0} \ .
\end{align}
For the mass deformation of the free scalar, we choose $\CO = \phi^2/2$ and $g_\CO = m^2$.

The partition function of a conformally coupled real massive scalar field on $\BS^1_\alpha\times \BH^{d-1}$ is given by \cite{Klebanov:2011uf}
\begin{align}\label{F_n_scalar}
        F_\alpha =  \int_0^\infty \d\lambda\, \mu_s (\lambda)\, \left[ \log\left( 1- e^{-2\pi \alpha \sqrt{\lambda+(mR)^2} }\right) + \pi \alpha \sqrt{\lambda+(mR)^2} \right] \ ,
\end{align}
where $\mu_s(\lambda)$ is the Plancherel measure of the real scalar field on $\BH^{d-1}$ of unit radius \cite{Camporesi:1990wm,Bytsenko:1994bc}\footnote{The volume $\Vol(\BH^{d-1})$ used in the Plancherel measure \eqref{Plancherel} is dimensionless.}
\begin{align}
\label{Plancherel}
        \mu_s (\lambda) = \frac{\Vol(\BH^{d-1})}{2^{d-1} \pi^{\frac{d+1}{2}} \Gamma\left(\frac{d-1}{2}\right)} \sinh (\pi \sqrt{\lambda})\left| \Gamma\left( \frac{d}{2} -1 + i \sqrt{\lambda}\right)\right|^2 \ .
\end{align}
This representation of the free energy leads to
\begin{align}\label{Fderiv}
	\lim_{\alpha\to 1} \partial_\alpha \partial_{(mR)^2} (F_\alpha - \alpha\, F_1)\big|_{(mR)^2 = 0} = -\frac{\pi^2}{2} \int_0^\infty \d\lambda\, \mu_s (\lambda)\,\frac{1}{\sinh^2 (\pi \sqrt{\lambda})} \ .
\end{align}

The $\lambda$ integral may be evaluated, making the change of variables $u^2 = \lambda$ and using the relation $u \sinh(\pi u) |\Gamma(i u)|^2 = \pi $ and the first Barnes lemma (see (D.1) of \cite{Smirnov:2006ry}),
 \begin{align}
 \label{firstBarneslemma}
\frac{1}{2 \pi } \int_{- \infty}^{ \infty} \d u \, \Gamma(\lambda_1 + iu) \Gamma(\lambda_2 + iu) &\Gamma(\lambda_3 -iu) \Gamma(\lambda_4-iu) \nonumber \\
&= \frac{\Gamma(\lambda_1 + \lambda_3) \Gamma(\lambda_1 + \lambda_4) \Gamma(\lambda_2 + \lambda_3) \Gamma(\lambda_2+\lambda_4)}{\Gamma(\lambda_1 + \lambda_2 + \lambda_3 + \lambda_4)} \ ,
 \end{align}
 with $\lambda_1 = \lambda_3 = 1$ and $\lambda_2 = \lambda_4 = d/2-1$.
 The answer is
 \be
\lim_{\alpha\to 1} \partial_\alpha \partial_{(mR)^2} (F_\alpha - \alpha\, F_1) |_{(mR)^2 = 0}  = -\Vol(\BH^{d-1}) \frac{(d-2) \Gamma(d/2 - 1)^2}{2^{d+2} \pi^{(d-3)/2} \Gamma((d+1)/2)}  \ .
\ee

 Next we need to evaluate $V_d$ which is the integral over the conformal factor \eqref{ConfFactor}
\begin{align}
 V_d =   \Vol(\BS^{d-2}) \int_0^{2\pi} \d \tau \int_0^\infty  \d u \,\sinh^{d-2} u\,\frac{\sin^2\theta_0}{(\cos \tau + \cos \theta_0\cosh u)^2 + \sin^2\theta_0\sinh^2 u} \ .
\end{align}
 A similar integral was carried out in appendix C of ref.\ \cite{Banerjee:2015tia}.  The $\tau$ integral can be done by contour integration, and after a change of variables $s= \tanh u$, the remaining integral can be put in the form
 \be
 V_d = 2 \pi  \Vol(\BS^{d-2}) \int_0^1 \d s \frac{s^{d-3}}{(1 - s^2)^{d/2-1}(s^2 \cot^2 \theta_0 + 1)} \ .
 \ee
 Where it converges, this integral reduces to\footnote{We used the volume $\Vol(\BS^{d-1}) = 2\pi^{d/2}/\Gamma \left( d/2\right)$.}
 \be
 V_d = - \frac{2 \pi^{(d+3)/2}  \sin^{d-2} \theta_0}{\sin(\pi d/2) \Gamma((d-1)/2)} \ .
 \ee
 Putting the pieces together, we find that
\begin{align}\label{s1_FE}
 s_1 = \frac{R^2 }{(d-2)^2} \frac{\pi^{3/2} \Gamma(d/2)^3 \csc(\pi d/2)}{4 \Gamma(d-2) \Gamma((d+1)/2)}  \sin^{d-2} \theta_0 \ .
\end{align}

For $d>3$, there is a divergence in the $s$ integral from the region near $s = 1$.  Instead of the dimensional regularization we just employed above, we may regulate the divergence by inserting a cut-off $\epsilon = 1 - s_{\rm max}$.  In the sphere to hyperbolic space mapping, the angular cut-off is related to the cut-off in $s$ via $\delta \theta^2 = 2 \epsilon \sin^2 \theta_0$.
It is useful to make a table of the results for small $d$ to see what is going on (see table \ref{tab:s1table}).
\begin{table}
\begin{align*}
\begin{array}{|c|c|}
\hline
d & \Delta S \\
\hline
3 & -\frac{\pi^3 (mR)^2}{32} \sin \theta_0 \\  &\\
4 & \frac{(mR)^2}{6} \sin^2 \theta_0 \log( \delta \theta)  + O(1) \\ & \\
5 &- \frac{3 \pi^2 (mR)^2}{1024} \sin^3 \theta_0 \left( \frac{2}{\delta \theta} - \pi  \right) \\ & \\
6 & - \frac{(mR)^2}{90} \sin^4 \theta_0 \left( \frac{1}{\delta \theta^2} + 2 \log \delta \theta + O(1) \right)\\
\hline
\end{array}
\end{align*}
\caption{Table of $\Delta S \equiv s_1 m^2$ for several $d$}
\label{tab:s1table}
\end{table}%
In even dimensions, the coefficient of the log is related to the result from dimensional regularization, while in odd dimensions, the $\delta \theta$ independent term is the result from dimensional regularization.

The result in $d=4$ dimensions can be written as an area law contribution
\be
\label{DeltaSd4}
\Delta S = \frac{1}{24 \pi} {\mathcal A}_\Sigma\, m^2 \log \epsilon \ ,
\ee
where $\Sigma = \BS^2$ is the entangling surface and ${\mathcal A}_{\Sigma} = 4 \pi R^2$ is its area.
This result matches the computation in ref.\  \cite{Hertzberg:2010uv} for a minimally coupled scalar.  More generally, as we review in section \ref{ss:universal}, the area law contribution scales as $m^{d-2} \log \epsilon$.  Thus only in $d=4$ is there an 
overlap region where we can compare the results.

In appendix \ref{sec:further}, we describe two additional methods for computing $s_1$ (\ref{s1init}).  Both methods rely on identifying the right hand side of eq.\,(\ref{s1init}) as the expectation value $\langle {:} \phi(x)^2 {:} \rangle_\alpha$ on the conical space $C_\alpha \times {\mathbb R}^{d-2}$.  The first method builds on the method of images employed by Cardy in ref.\ \cite{Cardy:2013nua} to compute the mutual information of a conformally coupled scalar in the limit where the two regions are far apart.  A key quantity in that paper was $\langle \phi(x) \phi(y)  \rangle_\alpha$.  Using some results from refs.\ \cite{Herzog:2014fra,Herzog:2014tfa}, it is straightforward to compute $\langle \phi(x) \phi(y)  \rangle_\alpha$ in the limit $x \to y$ and $\alpha \to 1$.  The second method computes $\langle {:} \phi(x)^2 {:} \rangle_\alpha$ directly from the path integral, assuming the boundary term (\ref{Ibryterm}) is present in the action.  This second method is remarkably efficient.

\subsection{Modular Hamiltonian method}

We now compare this free energy computation with the modular Hamiltonian method.  As noted at the beginning of this section, $\langle T_{\mu\nu}(x) {:} \phi(y)^2 {:} \rangle$ will vanish, and the contribution to $s_1$ comes entirely from the boundary term in the modular Hamiltonian (\ref{Hfixed}):
\begin{align}\label{S_shift}
	\begin{aligned}
		s_1 &= - \pi \xi  \int_{\partial\BH^{d-1}} \d^{d-2}x\,\sqrt{\gamma} \int_{\BR \times \BS^{d-1}} \d^d y\, \sqrt{g} \,\langle \phi^2(x) \phi^2(y)\rangle\ , \\
			&= - 2\pi \xi  \int_{\partial\BH^{d-1}} \d^{d-2}x\,\sqrt{\gamma} \int_{\BS^1 \times \BH^{d-1}} \d^d y\, \sqrt{g} \, e^{-2\sigma(y)}\,\langle \phi(x) \phi(y)\rangle^2 \ ,
	\end{aligned}
\end{align}
where we used the conformal map from $\BR\times \BS^{d-1}$ to $\BS^1 \times \BH^{d-1}$ and Wick's theorem in the second equality.
We will now derive this two-point function on $\BS^1 \times \BH^{d-1}$, that appears in the second line.

We derive the two-point function on  $\BS^1 \times \BH^{d-1}$ by using a Weyl transformation.
We start with the line element  \eqref{S1H2} on $\BS^1\times \BH^{d-1}$.
This metric is related to flat space with line element
\begin{align}
	\d s^2_{\BR^d} = \d t^2 + \d r^2 + r^2 \d\Omega_{d-2}^2 \ ,
\end{align}
via
\begin{align}
	\d s^2_{\BR^d} = \Omega^{-2} \,\d s^2_{\BS^1\times \BH^{d-1}} \ ,
\end{align}
where the Weyl factor is
\begin{align}\label{Weylfactor}
	\Omega = \cos\tau + \cosh u\ .
\end{align}
The explicit coordinate transformations are 
\begin{align}
	t = R\, \Omega^{-1}  \sin \tau \ , \qquad r = R\, \Omega^{-1} \sinh u \ .
\end{align}

The two-point function of a free scalar field of dimension $\Delta = (d-2)/2$ on $\BR^d$ takes the usual form
\begin{align}
	\langle \phi(x) \phi(y) \rangle_{\BR^d} = \frac{\CN}{|x-y|^{2\Delta}} \ ,
\end{align}
where we use a standard normalization, 
\be
\CN = \frac{1}{(d-2) \Vol(\BS^{d-1})} \ .
\ee
This normalization insures that the Laplacian acting on the two-point function gives a Dirac delta
function with unit strength.

The Weyl and coordinate transformations map this two-point function to\footnote{%
The invariant distance on $\BH^{d-1}$ is known to be
\begin{align*}
	d_{\BH^{d-1}} (u, \Omega_i ; u', \Omega_i') = \cosh u\, \cosh u' - \sinh u\, \sinh u'\, \sum_{i=1}^{d-1}\Omega_i\,\Omega_i' - 1 \ ,
\end{align*}
where $\Omega_i$ $(i=1,\cdots, d-1)$ are the embedding coordinates of $\BS^{d-2}$ satisfying $\sum_{i=1}^{d-1}\Omega_i^2 = 1$.
}

\begin{align}\label{GF_Hyp}
\begin{aligned}
	\langle \phi(u,\Omega_i) \phi(u',\Omega_i') \rangle_{\BS^1\times \BH^{d-1}} &= \Omega(x)^{-\Delta}  \Omega(y)^{-\Delta}  \langle \phi(x) \phi(y) \rangle_{\BR^d} \ ,\\
		&= \frac{\CN}{(2R^2)^\Delta \left(\cosh u\, \cosh u' - \sinh u\, \sinh u'  \sum_{i=1}^{d-1}\Omega_i\,\Omega_i' - \cos (\tau - \tau')\right)^\Delta} \ .
\end{aligned}
\end{align}

With the two-point function on $\BS^1\times \BH^{d-1}$ in hand, we can now evaluate the expression (\ref{S_shift}) for $s_1$.
We make use of the symmetry on $\BS^1\times \BH^{d-1}$ to fix the point $x \in \partial \BH^{d-1} = \BS^{d-2}$ to the ``north pole'', $\Omega_i = \delta_{i1}$.  This choice significantly simplifies the calculation.
Plugging \eqref{GF_Hyp} into \eqref{S_shift} with \eqref{ConfFactor}, we get
\begin{align}\label{ShiftEE}
	\begin{aligned}
		s_1 &= -2\pi \xi\, \Vol(\BS^{d-2})\,R^{2d-2}\,\sinh^{d-2} u\\
			&\qquad\qquad \cdot \int_{\BS^1\times \BH^{d-1}} \sinh^{d-2}u'\,\d u'\,\d\tau'\,\d\Omega_{d-2}' \frac{\sin^2\theta_0}{(\cos\tau' + \cos\theta_0\,\cosh u')^2 + \sin^2\theta_0\,\sinh^2 u'}\\
			&\qquad \qquad\quad \cdot \frac{\CN^2}{(2R^2)^{d-2} \left(\cosh u\, \cosh u' - \sinh u\, \sinh u'  \cos\theta' - \cos (\tau - \tau')\right)^{d-2}}\Bigg|_{u\to \infty} \ ,\\
			 &= 
			 -\frac{R^2}{(d-2)^2}\frac{\pi^{\frac{1}{2}} \Gamma(d/2)^3}{2 \Gamma(d-2) \Gamma((d+1)/2)}
			 \int_0^1 \d s\, \frac{s^{d-3}}{(1-s^2)^{d/2-1}(s^2\cot^2\theta_0+1)} \ .
	\end{aligned}
\end{align}
The $\tau'$ integral can be done by contour integration.  The $s$-integral is related to the $u'$-integral by the change of variables $s = \tanh(u')$.
This expression is identical to what we obtained using the free energy, \eqref{s1_FE}.

We have also computed these mass corrections on the sphere by conformally mapping them to a conical space rather than hyperbolic space.  The calculation is far lengthier, and involves Feynman parameters.  At the end of the day, the relation between the UV cut-offs in the two calculations is obscured.  However, we can still match the universal terms.  We will not include the details of this calculation.

\subsection{Lattice calculation}
\label{sec:lattice}

\begin{figure}[h]
\begin{center}
	\begin{subfigure}[b]{0.45\textwidth}
		\includegraphics[width=2.8in]{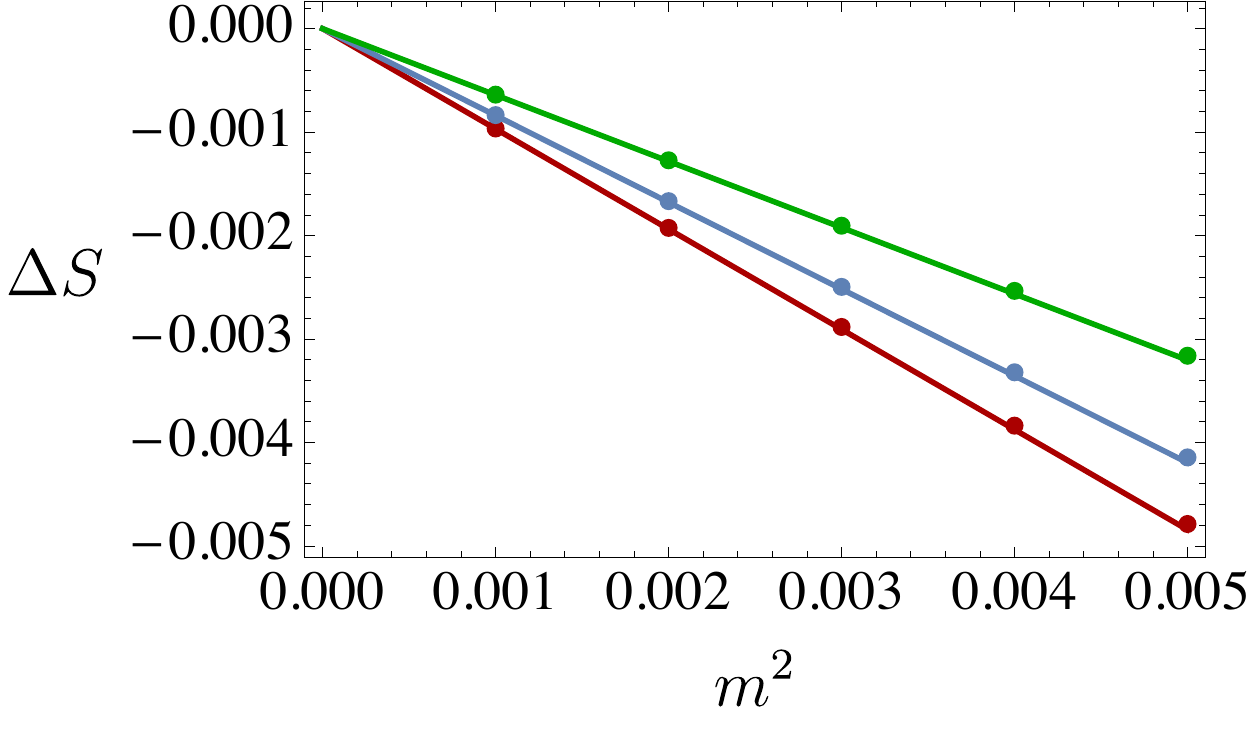}
		\caption{Linearity in $m^2$}
	\end{subfigure}
	~~
	\begin{subfigure}[b]{0.45\textwidth}
		\includegraphics[width=3.2in]{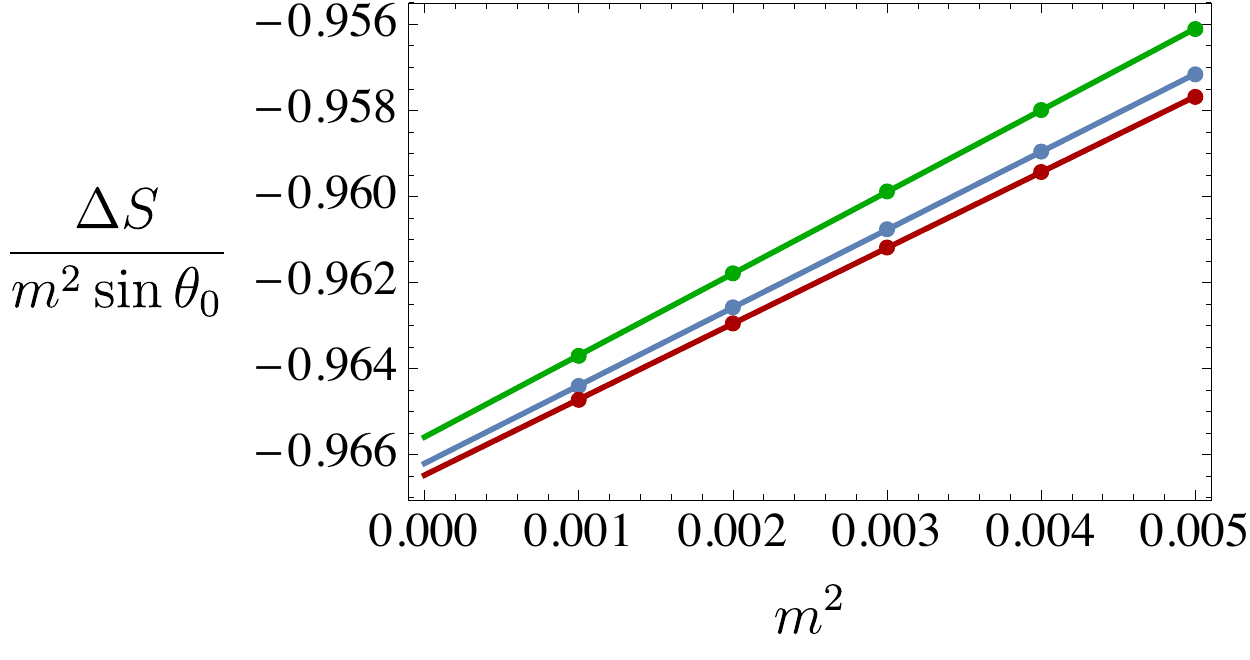}
		\caption{Angular dependence}
	\end{subfigure}
	~
	\begin{subfigure}[b]{0.45\textwidth}
		\includegraphics[width=3in]{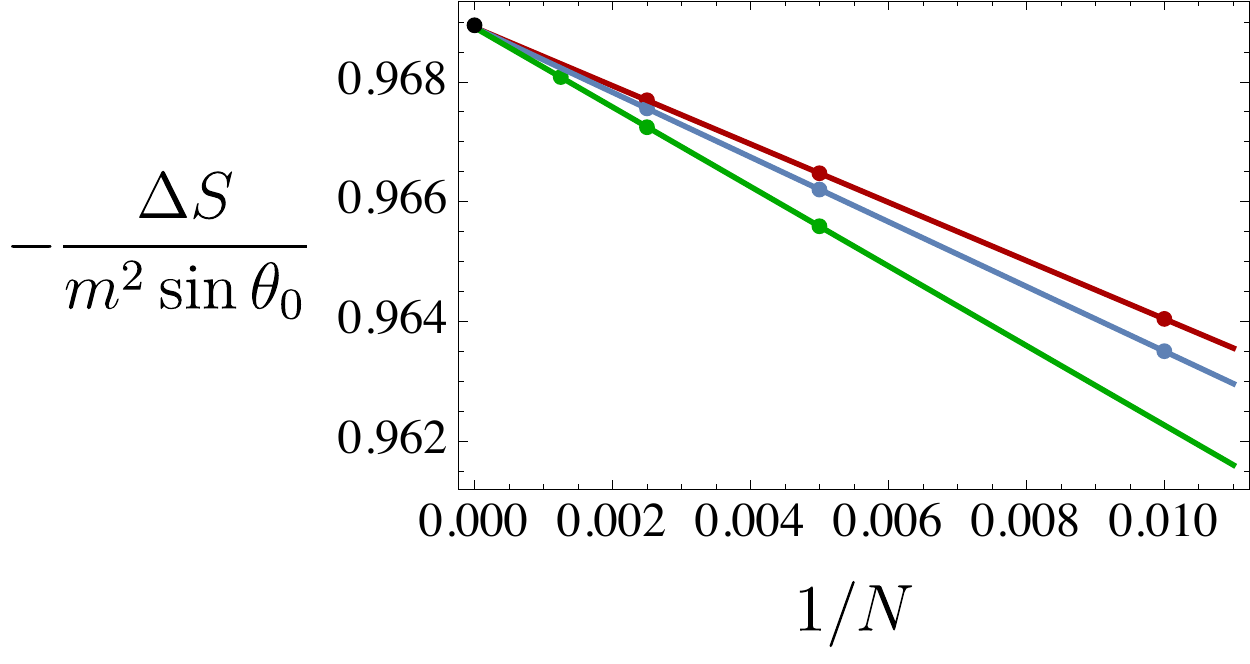}
		\caption{Lattice spacing dependence}
	\end{subfigure}
\end{center}
\caption{
Entanglement entropy plots for the massive scalar in $d=3$.  The data points are numerically determined while the straight lines are fits.  From top to bottom in (a) and (b) and from bottom to top in (c), the three lines in each graph correspond to $\theta_0 = 41.4^\circ$, $60^\circ$ and $90^\circ$.  The mass is measured in units where the radius of the sphere $R=1$.  In plots (a) and (b), a grid of 200 points in $\cos \theta$ was used to discretize the sphere, where $\theta$ is the polar angle, while in (c), the number of grid points was varied. 
Panel (a) shows the linearity in $m^2$ of the change in entanglement entropy.
Panel (b) demonstrates that the angular dependence of the change in entropy is well approximated by $\sin \theta_0$, as predicted.  Additionally, the linearity indicates that the next higher order correction to the change in entropy scales as $m^4$.
In panel (c), the $x$-axis is the inverse of the number of grid points on the sphere.  The data points were computed from the $y$-intercept of fits like in panel (b), but for a variety of different grid sizes.  The black dot corresponds to the expected coefficient $\pi^3/32$ of the change in entanglement entropy in the continuum limit.  The error in the linear extrapolation is about 4 parts in $10^5$.    
 \label{fig:masspert3}
}
\end{figure}

As a third independent check, we calculate the entanglement entropy of a massive scalar numerically, using a variant of Srednicki's method \cite{Srednicki:1993im}.  Further details of our approach can be found in appendix \ref{sec:numerics}. 
In brief, we begin with the canonical Hamiltonian for a massive scalar on a sphere $\BS^{d-1}$.  As the sphere has no boundary, issues about codimension one and two boundary terms will not arise.  We decompose this Hamiltonian into spherical harmonics on $\BS^{d-2}$, leaving an overall dependence on the polar angle $\theta$ of $\BS^{d-1}$.   We then compute discretized two-point functions of the field $\phi$ and conjugate momentum $\Pi$ for each angular momentum mode, as a function of the polar angle $\theta$.  Given the Gaussian nature of the ground state wave function, from these two-point functions, restricted to a cap on the sphere with opening angle $2 \theta_0$, we can reconstruct the eigenvalues of the reduced density matrix and thus compute the entanglement entropy. 

Given the results of the previous subsections, there is an obvious issue with UV divergences for all $d \geq 4$. 
Only in $d=3$ is the check straightforward.
In figure \ref{fig:masspert3}a, we see that the corrections to the entanglement entropy are indeed linear in $m^2$ for small $m^2$.  A more careful analysis of the data, shown in figure \ref{fig:masspert3}b, reveals that the linear term has a $\sin \theta_0$ dependence, and that the corrections to the linear scaling are $O(m^4)$, in agreement with the expectation from conformal perturbation theory.
We can do even better and try to compute the $\pi^3/32$ coefficient of the entanglement change numerically.  
Computing the entanglement entropy for several different grid sizes and extrapolating to get a continuum limit, we are able to obtain the numerical
value $\pi^3/32 = 0.968946$ to about four parts in $10^5$.  The results are shown in figure \ref{fig:masspert3}c.

\begin{figure}[h]
\begin{center}
	\begin{subfigure}[b]{0.45\textwidth}
		\includegraphics[width=2.8in]{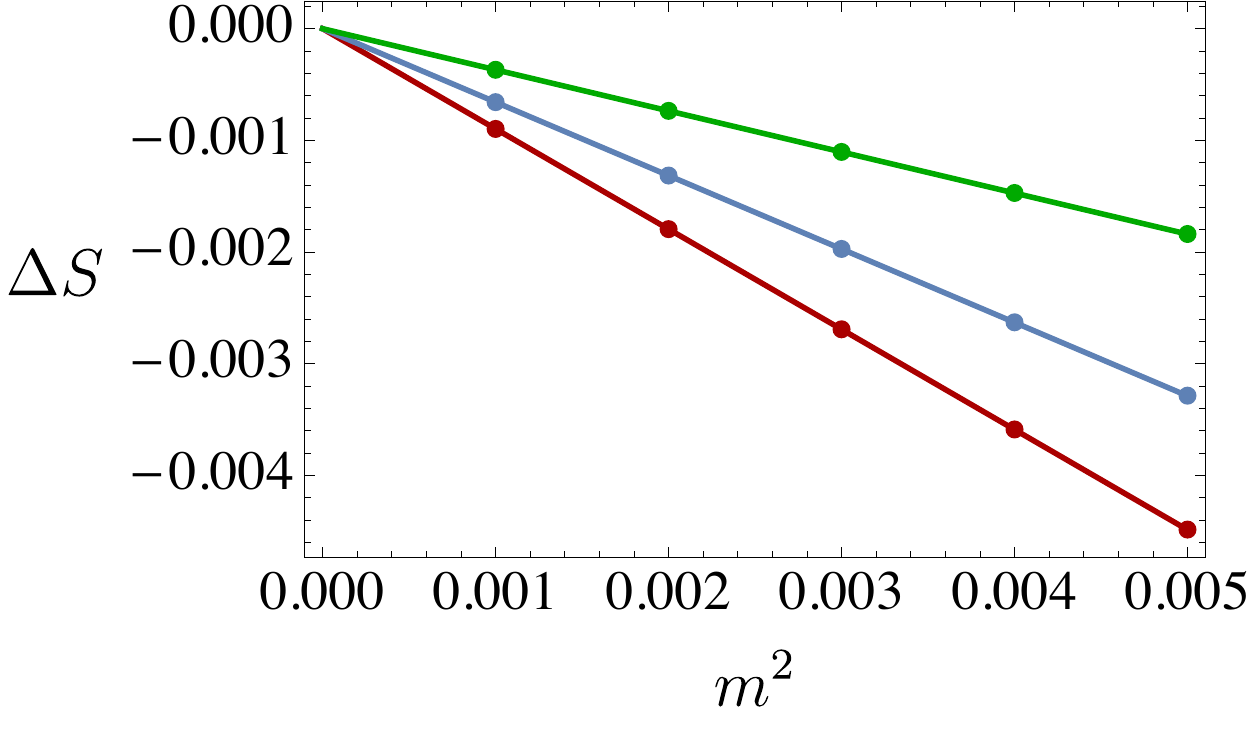}
		\caption{Linearity in $m^2$}
	\end{subfigure}
	~
	\begin{subfigure}[b]{0.45\textwidth}
		\includegraphics[width=3.1in]{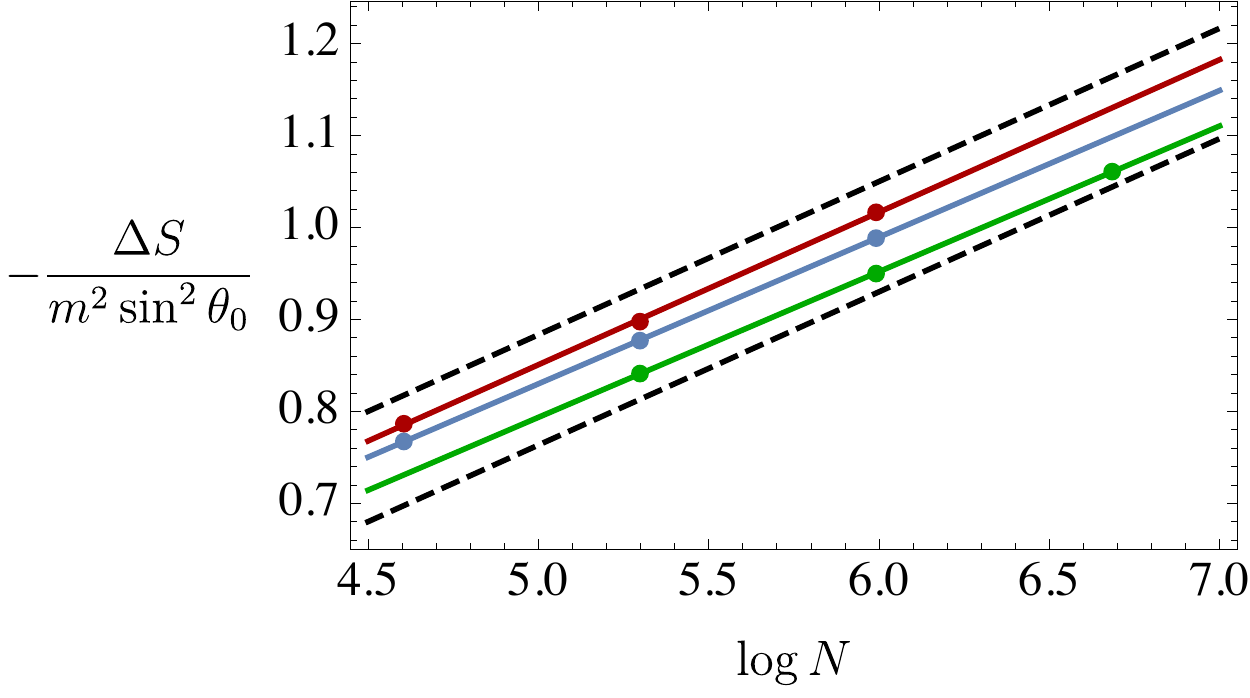}
		\caption{Lattice spacing dependence}
	\end{subfigure}
\end{center}
\caption{Entanglement entropy plots for the massive scalar in $d=4$.  Different lines correspond to 
different opening angles $2 \theta_0$.  From top to bottom in (a) and bottom to top in (b), $\theta_0 = 41.4^\circ$, $60^\circ$ and $90^\circ$.  The mass is measured in units where the radius of the sphere $R=1$.   The points are numerically computed, and the lines are linear fits.
Panel (a) demonstrates the linear dependence in $m^2$ of the entanglement entropy correction.   A grid of 200 points in $\cos \theta$ was used to discretize the sphere, where $\theta$ is the polar angle.   In panel (b), $N$ is the number of points used to discretize the sphere.  The data points were obtained from fits to data like in panel (a), but varying the grid size.  The dashed lines with the expected slope of 1/6 are a guide to the eye.  The slope of the fits are a bit low, and range from bottom to top, from 0.158 to 0.1659.
\label{fig:masspert4}
}
\end{figure}

Already in $d=4$, there is a log dependence on the cut-off which makes extracting precise results from the numerics more involved.  
Figure \ref{fig:masspert4}a indicates that the corrections are still linear in $m^2$, as predicted.  Figure \ref{fig:masspert4}b shows that the corrections additionally obey a $\sin^2 \theta_0$ dependence, as predicted by table \ref{tab:s1table}.  Having varied the lattice spacing, we also get a rough fit for the coefficient of the $\log$.  The answer depends slightly on $\theta_0$.  For $\theta_0 = \pi/2$, we get the best fit of $0.1659$ which is very close to the analytic prediction of 1/6.  For $\theta=41.4^\circ$, we get the worst fit value of $0.158$.  

\section{Large Mass Expansion}
\label{sec:largemass}

Entanglement entropy of an entangling (closed) curve $\Sigma$ in a gapped $(2+1)$-dimensional system is expected to have a large gap expansion 
\begin{align}\label{Large_mass_exp}
	S_\Sigma = \alpha \frac{\ell_\Sigma}{\epsilon} + \beta\,m\,\ell_\Sigma - \gamma + \sum_{n=0}^\infty \frac{c^\Sigma_{2n+1}}{m^{2n+1}} \ ,
\end{align}
where $m$ is a gap scale and $\ell_\Sigma$ is the length of $\Sigma$ \cite{Grover:2011fa}. 
An algorithm to fix the coefficients $c^\Sigma_{2n+1}$ for free fields was proposed in \cite{Klebanov:2012yf}, which relates $c^\Sigma_{2n+1}$ to a conformal anomaly in a $(2n+4)$-dimensional theory compactified on $\BT^{2n+1}$.
The authors of ref.\ \cite{Klebanov:2012yf} focused on the flat space case ${\mathbb R}^{2,1}$.  Attempting to extend the proposal to the cylinder ${\mathbb R} \times \BS^2$, ref.\  \cite{Banerjee:2015tia} found a small discrepancy for conformally coupled scalars.  The purpose of this section is to explain and remove the discrepancy.  The resolution has no direct connection to the boundary terms that form the main topic of this paper; it is instead related to the dimension dependence of the conformal coupling parameter $\xi$.

Nevertheless, we will be able to perform an interesting cross-check as a result of the computation here.  The value 
\be
\label{betavalue}
\beta = - \frac{1}{12} \ ,
\ee
was established for a minimally coupled scalar field in refs.\ \cite{Hertzberg:2010uv,Huerta:2011qi}.  One interpretation of the results here is as evidence that $\beta$ remains equal to $-1/12$ in the presence of a non-minimal coupling $\xi$ to curvature, consistent with our discussion of area law contributions in the next section.

We begin by reviewing how $c^\Sigma_1$ can be determined for a free massive scalar on ${\mathbb R}^{2,1}$. 
We start with a free massless scalar field in $3+1$ dimensions and compactify a spatial direction to a circle of circumference $L$. 
Then the theory reduces to an infinite tower of free massive scalars in $2+1$ dimensions with masses 
\begin{align}\label{KKmass}
	m^2_n = \left( \frac{2\pi}{L}\right)^2 n^2 \ , \qquad n\in \BZ \ .
\end{align}
If the entangling surface in $3+1$ dimensions is topologically a torus $\Sigma_2 = \Sigma \times \BS^1$, the entanglement entropy $S^{(3+1)}_{\Sigma_2}$ becomes a sum over the entropies $S^{(2+1)}_\Sigma (m_n)$ across the entangling curve $\Sigma$ for the massive free scalars:
\begin{align}
	S^{(3+1)}_{\Sigma_2} = \sum_{n\in \BZ} S^{(2+1)}_\Sigma (m_n) \ .
\end{align}
Taking the $L\to \infty$ limit, the Kaluza-Klein mass becomes continuous $(m_n \to p)$ and the entropy turns out to be
\begin{align}
\label{intrel}
	S^{(3+1)}_{\Sigma_2} = \frac{L}{2\pi}\cdot 2 \int_0^{1/\epsilon} \d p\,S^{(2+1)}_\Sigma (p) \ ,
\end{align}
where we introduced a UV cutoff $\epsilon$ for the KK mass.
On the left hand side, there is a logarithmic divergence $s_0\log \epsilon$ whose coefficient $s_0$ is fixed by a conformal anomaly, while the corresponding term arises from the order $1/m$ term in \eqref{Large_mass_exp}.
Matching these terms, we find 
\begin{align}\label{c1}
	c^\Sigma_1 = - \frac{\pi}{L} s_0 \ .
\end{align}

Given an entangling surface in CFT$_4$, $s_0$ can be fixed by Solodukhin's formula \cite{Solodukhin:2008dh,Fursaev:2013fta} 
\begin{align}
\label{Solodukhin}
	s_0^\text{Solodukhin} = \frac{a}{2}\,\chi[\Sigma_2] + \frac{c}{2\pi} \int_{\Sigma_2}\left[ \CR_{aa} - \CR_{abab} - \frac{\CR}{3} + k_{\mu\nu}^a k_a^{\mu\nu} - \frac{1}{2}(k^{a\,\mu}_\mu)^2\right] \ ,
\end{align}
where $a$ and $c$ are theory dependent central charges, chosen here for the scalar theory, and normalized such that $(a,c) = (\frac{1}{180}, \frac{1}{120})$ respectively.
$\chi[\Sigma_2]$ is the Euler characteristic of $\Sigma_2$, $\CR$ the Ricci scalar, and $\CR_{aa} \equiv \sum_{a}\CR^{\mu\nu}n^a_\mu n^a_\nu$, $\CR_{abab} \equiv \sum_{a,b}\CR^{\mu\nu\rho\sigma}n^a_\mu n^b_\nu n^a_\rho n^b_\sigma$ are projected Riemann tensors by the normal vectors $n^a_\mu$ $(a=1,2)$ to $\Sigma_2$.
$k^a_{\mu\nu} \equiv \gamma_\mu^{\,\rho}\gamma_\nu^{\,\sigma}\nabla_\rho n^a_\sigma$ is the extrinsic curvature with the induced metric $\gamma_{\mu\nu}\equiv g_{\mu\nu} - \sum_a n^a_\mu n^a_\nu$.
Applying the formula to the entangling surface $\Sigma$ that is topologically a torus $\chi[\Sigma_2]=0$ and reducing the theory on $\BS^1$, one can determine $c^\Sigma_1$ as a function of the extrinsic curvature of the entangling curve $\Sigma$ on a plane, which is consistent with a numerical calculation \cite{Klebanov:2012yf}.

Superficially, nothing about the above argument seems restricted to ${\mathbb R}^{2,1}$.  We may try to use Solodukhin's formula (\ref{Solodukhin}) and the relation (\ref{c1}) to determine $c^\Sigma_1$ for an arbitrary entangling surface $\Sigma$ in an arbitrary two-dimensional manifold $M_2$ for a spacetime of the form ${\mathbb R}^1 \times M_2$.  Ref.\ \cite{Banerjee:2015tia} made just such an attempt when $M_2 = {\mathbb S}^2$.  
Parametrizing a curve $\Sigma$ by $\theta = \Theta (\phi)$ on $\BS^2$ with line element $\d s^2_{\BS^2} =R^2( \d\theta^2 + \sin^2 \theta \d\phi^2)$, 
one finds 
\begin{align}\label{c1formula}
	c_1^\Sigma = - \frac{c}{2}\int_\Sigma \left[ \frac{1}{3R^2} + \kappa^2 \right] \ ,
\end{align}
where the extrinsic curvature $\kappa$ of $\Sigma$ is
\begin{align}
	\kappa^2 \equiv k_{\mu\nu}^a k_a^{\mu\nu} - \frac{1}{2}(k^{a\,\mu}_\mu)^2 = \frac{\left[ 2\cos\Theta\, \Theta'^2 + \sin\Theta (\sin\Theta\cos\Theta - \Theta'')\right]^2}{2R^2 (\sin^2 \Theta + \Theta'^2)^3} \ .
\end{align}
(We used the curvatures $\CR = 2/R^2$, $\CR_{aa} = 1/R^2$ and $\CR_{abab}=0$.)

\begin{figure}
	\centering
	\includegraphics[width=4in]{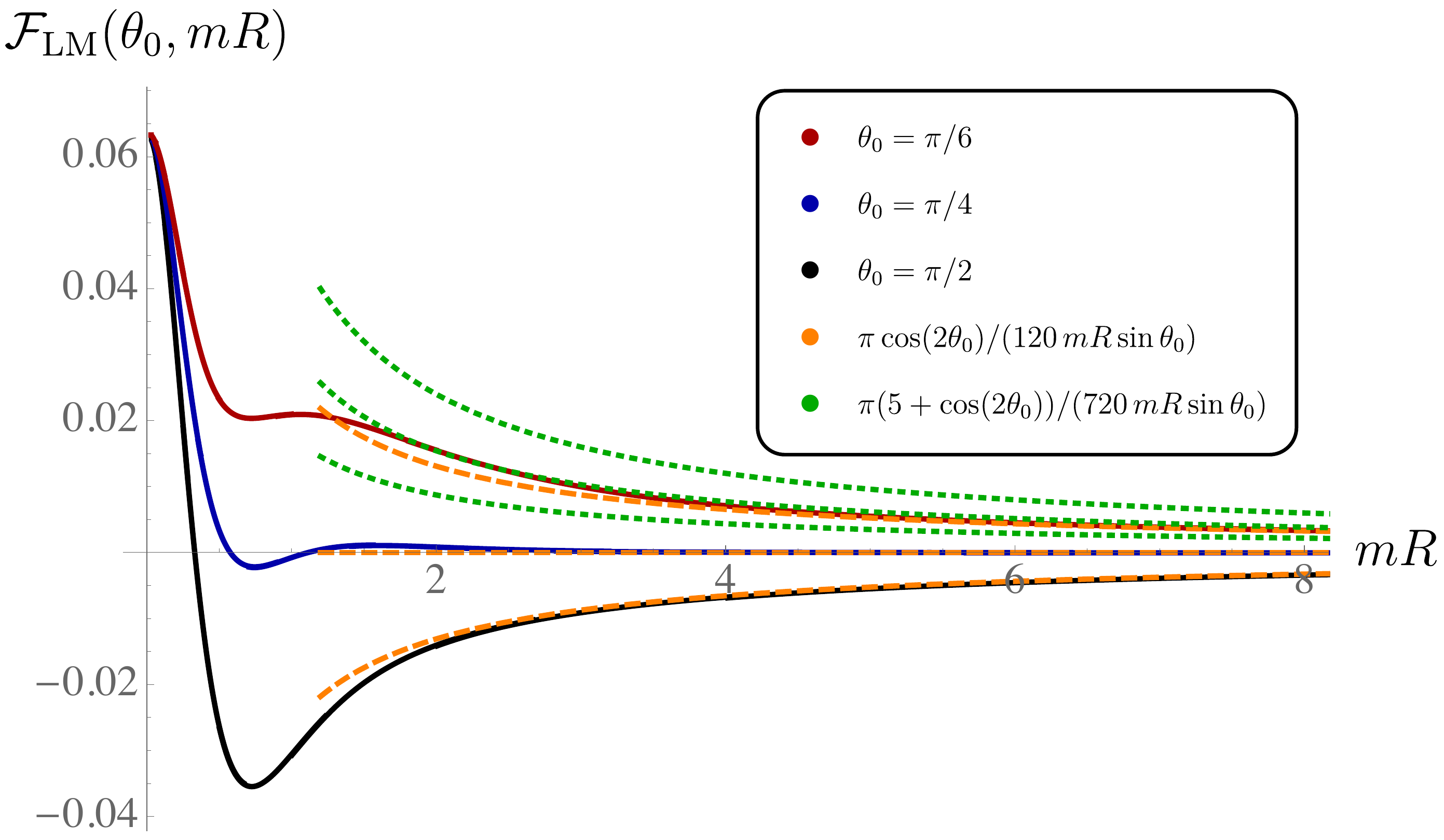}
	\caption{Plots of the Liu-Mezei type renormalized entanglement entropy $\CF_\text{LM}(\theta_0, mR) \equiv (R \partial_R -1)S (\theta_0, mR)$ for a free massive scalar field. From top to bottom, the three curves correspond to entangling regions with opening angles $2\theta_0 = \pi/3, \pi/2$ and $\pi$, respectively.
	The green dotted lines represent the leading large mass expansions calculated with the coefficient $c_1^\text{cap}$ in \eqref{c1cap}, while the orange dashed lines take into account the shift $\delta c_1^\text{cap}$  \eqref{c1shift} to the coefficient and show better fits to the numerical results.}
	\label{fig:LargeMassPlot}
\end{figure}

To make a numerical check, it is useful to restrict to the simple case $\Theta (\phi) = \theta_0$ of a cap-like entangling region with opening angle $2\theta_0$.
The coefficient \eqref{c1formula} for a free scalar simplifies to
\begin{align}\label{c1cap}
	c_1^\text{cap} = - \frac{\pi}{120 R}\sin\theta_0 \left( \frac{1}{3} + \frac{\cot^2\theta_0}{2}\right) \ .
\end{align}
However, a discrepancy was found between the analytic and numerical calculations in the large mass expansion \cite{Banerjee:2015tia}.
The numerics suggests an additional contribution to the coefficient (see figure \ref{fig:LargeMassPlot})
\begin{align}\label{c1shift}
	\delta c_1^\text{cap} = \frac{\pi \sin\theta_0}{144R} \ .
\end{align}

\subsection{Resolution by conformal coupling}

The issue is that in fact the conformal coupling cannot be ignored on a general spatial manifold $M_2$.
The discrepancy comes from the dimension dependence of the conformal coupling.  
  In flat space, the mass contribution from the conformal coupling is zero, and there is no discrepancy.  On the sphere, however, the dimension dependence leads to mixing between the $O(m)$ and $O(1/m)$ terms in the expansion (\ref{Large_mass_exp}).  
The essential point is that in the expansion (\ref{Large_mass_exp}), the mass gap $m$ is defined with reference to the three-dimensional coupling $\xi_3 = 1/8$ while in the integral (\ref{intrel}), the mass $p$ should be defined with respect to the coupling $\xi_4 = 1/6$.  
We have in general that
\be
m^2 + \xi_3 \,\CR^{(2+1)} = p^2 + \xi_4\,\CR^{(3+1)} \ .
\ee
Given this relation, we can rewrite two terms in the large $m$ expansion in terms of $p$:
\be\label{LMrelation}
\beta\, m\, \ell_\Sigma + \frac{c_1^\Sigma}{m} = \beta\, p\, \ell_\Sigma + \frac{1}{p} \left[ c_1^\Sigma + \frac{1}{2}\left(\xi_4\,\CR^{(3+1)} -\xi_3 \,\CR^{(2+1)}\right)  \beta\, \ell_\Sigma  \right] + O(p^{-3})\ .
\ee
In our setup, $\CR^{(3+1)} = \CR^{(2+1)} = \CR(\BS^2) = 2/R^2$, $\ell_{\Sigma} =2\pi R \sin\theta_0$, while 
$\beta = -1/12$ for a free scalar field \cite{Hertzberg:2010uv,Huerta:2011qi}.  Assembling the pieces, we find that the second term in the square bracket in \eqref{LMrelation} agrees with \eqref{c1shift}
and resolves the discrepancy found in \cite{Banerjee:2015tia}.  That a value of $\beta = -1/12$ is required for the discrepancy to be resolved can be viewed as evidence that the value of $\beta$ is independent of the conformal coupling $\xi$.

To show more clearly how much the shift term \eqref{c1shift} improves fitting in the large mass region, we plot the numerical and analytic results for a few choices of $\theta_0$ in figure \ref{fig:LargeMassPlot}.
We introduced the renormalized entanglement entropy of Liu-Mezei type \cite{Liu:2012eea}
\begin{align}
	\CF_\text{LM}(\theta_0, mR) \equiv (R\partial_R - 1)S (\theta_0, mR) \ ,
\end{align}
to remove the UV divergence.\footnote{This renormalized entanglement entropy does not have a monotonicity property under the RG flow in contrast to the flat space case \cite{Casini:2012ei}.
Another type of renormalized entanglement entropy $\CF_\text{C} \equiv (\tan\theta_0 -1)S(\theta_0)$ on the cylinder is proposed and shown to be monotonic numerically for a free massive scalar field in \cite{Banerjee:2015tia}.
}
The leading large mass expansions with and without the shift term \eqref{c1shift} are shown in the orange dashed and green dotted lines respectively.
The orange lines fit the numerical data beautifully even in the $mR \approx 2$ region.

\subsection{Alternate resolution}
\label{sec:largemassalt}

An alternative way to fix the large mass discrepancy is to reduce a free massless scalar with non-conformal coupling $\xi_3$ in four dimensions to an infinite tower of the KK modes with masses $m = m_n$ \eqref{KKmass} in three dimensions.
Namely, we uplift the theory of a free massive scalar with conformal coupling in $2+1$ dimensions to a free scalar theory in $3+1$ dimensions with the action
\begin{align}\label{4d_action}
	\action^{(3+1)}_E = \frac{1}{2}\int_{\CM \times \BS^1} \d^4x\, \sqrt{g}\left[ (\partial\phi)^2 + \xi_3\, \CR^{(3+1)}\phi^2 \right] \ .
\end{align}
Note that this action is not conformally invariant as the coefficient of the curvature coupling differs from the correct value $\xi_4 = 1/6$.
A manifold such as a cylinder $\CM=\BR\times \BS^2$ has a constant Ricci scalar, and the conformal mass can be treated on the same footing as a usual mass.
As we discussed already around eq.\ (\ref{DeltaSd4}) and as we will discuss in more detail in section \ref{ss:universal}, one can show that the logarithmic divergence of the entropy for the theory \eqref{4d_action} has a contribution additional to Solodukhin's formula:
\begin{align}
	s_0 = s_0^\text{Solodukhin} + \delta s_0 \ ,
\end{align}
where $\delta s_0$ is a universal area term.
If one naively regards the action \eqref{4d_action} as a conformally coupled scalar with the mass $\left(\xi_3- \xi_4\right)\,\CR^{(3+1)}$, one would get from eq.\ \eqref{DeltaSd4} 
\begin{align}\label{HWformula}
	\delta s_0 = \frac{1}{24\pi}\left(\xi_3 - \frac{1}{6}\right)\CA_{\Sigma_2} \CR^{(3+1)} \ ,
\end{align}
where $\CA_{\Sigma_2}$ is the area of the entangling surface $\Sigma_2$.
In our setup, we find
\begin{align}
	\delta s_0 = - \frac{L\sin\theta_0}{144 R} \ ,
\end{align}
which yields a shift for $c_1^\Sigma$ given by \eqref{c1shift} using the relation \eqref{c1}.

Though the discrepancy has been successfully resolved, we emphasize that the formula \eqref{HWformula} only works for a manifold with a constant curvature.
For a general manifold, ref.\ \cite{Lewkowycz:2012qr} argues a formula based on several examples 
\begin{align}
	\delta s_0^\text{LMS} = \frac{1}{24\pi}\left(\xi_3 - \frac{1}{6}\right) \int_{\Sigma_2} \d^2 \sigma \sqrt{h}\, \CR (h) = \frac{1}{6} \left(\xi_3 - \frac{1}{6}\right) \chi[\Sigma_2]\ ,
\end{align}
where $h$ is the induced metric on the entangling surface $\Sigma_2$, and $\chi[\Sigma_2]$ is the Euler number.
If this formula were true, we would find $\delta s_0 = 0$ as our entangling surface is topologically a torus and come back to the discrepancy.
Instead we propose an alternative formula
\begin{align}\label{ourFormula}
	\delta s_0^\text{our} = \frac{1}{24\pi}\left(\xi_3 - \frac{1}{6}\right) \int_{\Sigma_2} \d^2 \sigma \sqrt{h}\, \CR (g)\ ,
\end{align}
with the Ricci curvature $\CR(g)$ of the background metric.
Our formula differs from theirs by the extrinsic curvature terms due to the Gauss-Codazzi relation, and thus correctly reproduces the result of ref.\ \cite{Lewkowycz:2012qr} for the spherical waveguide geometry.
Our proposal \eqref{ourFormula} may be testable by putting the theory on an ellipsoid instead of $\BS^2$ to see the dependence on the background curvature.

\section{Universal Area Term}\label{ss:universal}

In a field theory with a mass gap $m$, the entanglement entropy is believed to obey a universal area law for any shape of an entangling surface on flat spacetime \cite{Hertzberg:2010uv}:
\begin{align}
	m^2\partial_{m^2} S = \gamma_d\, \CA_\Sigma 
				\begin{cases}
					m^{d-2} \log (m/\epsilon) 	\qquad & (d:\text{even}) \ , \\
					m^{d-2}	& (d:\text{odd}) \ ,
				\end{cases} 
\end{align}
where $\gamma_d$ is a dimension-dependent constant and $\CA_\Sigma$ is the area of the entangling surface.
Employing the heat kernel method on the replica manifold, the constants $\gamma_d$ were determined for free massive scalar and fermion fields \cite{Hertzberg:2010uv} with minimal couplings to the background geometry.
There are, however, a one-parameter family of non-minimal couplings $\xi\,\CR \phi^2$ for a scalar field when coupled to a curved space and it is not so obvious whether the universal area term depends on the coupling $\xi$.
Ref.\ \cite{Casini:2014yca} argues the minimal choice to be natural on physical grounds as the operator algebra on flat space does not depend on $\xi$.

The area law dependence can be derived both from the Adler-Zee formula \cite{Adler:1980pg,Zee:1980sj,Adler:1982ri} and from the Rosenhaus-Smolkin formula \cite{Rosenhaus:2014nha,Rosenhaus:2014ula}.  With some assumptions, ref.\ \cite{Casini:2014yca} explicitly demonstrates these two methods are equivalent.  Depending on the precise treatment of boundary terms, contact terms and the equations of motion, both methods can also produce a dependence of $\gamma_d$ on $\xi$.   

In what follows we revisit and address the $\xi$ dependence of $\gamma_d$ from the perspective of the discussion in section \ref{sec:bryterm}.

\subsection{Adler-Zee formula}
In general, the effective action induced by matter coupled to gravity has a derivative expansion of the metric, respecting diffeomorphism invariance:
\begin{align}
	I_\text{eff} = \int \d^d x\, \sqrt{g}\, \left[ c_0 + c_1 \CR + O(\CR^2) \right] \ .
\end{align}
The first few expansion coefficients are completely determined by flat space expectation values of the stress tensor as \cite{Adler:1980pg,Zee:1980sj,Adler:1982ri,Muratani:1983hh}
\begin{align}
	\begin{aligned}
		c_0 &= - \frac{\langle \Theta\rangle}{d} \ , \\
		c_1 &= \frac{1}{4d(d-1)(d-2)}\int \d^d x\, {\vec x}^{\,2}\, \langle \Theta(x)\Theta(0)\rangle -\frac{1}{d-2} \biggl\langle \frac{\delta \Theta}{\delta\CR}\biggr\rangle	\ ,
	\end{aligned}
\end{align}
where $\Theta = T_\mu^{~\mu}$ and the variation of the stress tensor in the second term of $c_1$ is taken for a conformally flat metric.
The expression for $c_1$ is called the Adler-Zee formula, and interpreted by them as a renormalization of Newton's constant.

A familiar process of evaluating the effective action on a conical deficit around the entangling surface $\Sigma$ yields the leading area term of the entanglement entropy 
\begin{align}
\label{entropyresult}
	S = - 4\pi c_1\CA_\Sigma + \cdots \ .
\end{align}
This is a universal area formula valid for any theory defined on any space, but the coefficient $c_1$ appears to depend on the type of the stress tensor to be used.
For the scalar field, in view of the discussion in section \ref{sec:pathintegral}, we should use the stress tensor derived in flat space from the minimally coupled Euclidean action (\ref{actionE}).  This action of course does not depend on $\xi$ and neither as a result will the area contribution to the entropy.

To compare with the literature, we consider a two parameter space of stress tensors.  One parameter is associated with the non-minimal coupling $\xi$ in the improvement term.  The other is associated with the possibility of using the equations of motion.  
For a non-minimally coupled scalar with mass, the trace of the stress tensor is
\begin{align}\label{Theta}
	\Theta = - \frac{d-2}{2}(\partial\phi)^2 + \xi (d-1)\nabla^2 \phi^2 - \frac{d-2}{2} \xi\, \CR\,\phi^2 - \frac{d}{2} m^2 \phi^2 \ .
\end{align}
We can further modify the trace using the equations of motion
\begin{align}
	\Theta_{(\kappa)} = \Theta(x) + \frac{d-2}{2}\kappa\,\phi(-\nabla^2 + \xi\,\CR + m^2)\phi \ ,
\end{align}
where $\kappa$ can be arbitrary.  The value $\kappa=\xi / \xi_c$ is particularly nice because it removes the $\phi \nabla^2 \phi$ dependence from the trace in general and in the conformally coupled case ($\xi = \xi_c$ and $m=0$) sets $\Theta_{(1)} = 0$.

The integral of the correlator of $\Theta$ on flat space is performed with Wick contraction
\begin{align}
\begin{aligned}
	\int \d^d x\, {\vec x}^{\,2}\, \langle \Theta(x)\Theta(0)\rangle &= \int \d^d x\, {\vec x}^{\,2}\int \frac{\d^d p}{(2\pi)^d}\frac{\d^d q}{(2\pi)^d} e^{i(p+q)x}\frac{2\left( \frac{d-2}{2}p\cdot q - (d-1)\xi (p+q)^2 - \frac{d}{2}m^2\right)^2}{(p^2+m^2)(q^2+m^2)}  \ ,\\
		&= -\int\frac{\d^d p}{(2\pi)^d}\, \partial_p \partial_q \frac{2\left( \frac{d-2}{2}p\cdot q - (d-1)\xi (p+q)^2 - \frac{d}{2}m^2\right)^2}{(p^2+m^2)(q^2+m^2)} \bigg|_{q=-p} \ , \\
		&= \int\frac{\d^d p}{(2\pi)^d}\, \left[ -\frac{8m^6}{(p^2+m^2)^4} + \frac{8m^4}{(p^2+m^2)^3} + \frac{(3d - 2)(d -2)m^2}{(p^2+m^2)^2}\right. \\
		&\qquad\qquad\qquad \left.+ \frac{(d-1)(d-2)^2}{p^2+m^2}
		- 4(d-1)\xi\, \partial_{p_\mu} \frac{p^\mu}{p^2 + m^2}\right] \ , \\
		&= - \frac{d(d-1)(d-2)m^{d-2}}{3\cdot 2^d \pi^{\frac{d}{2}-1}\sin\left(\frac{\pi d}{2}\right) \Gamma(d/2)} \ ,
\end{aligned}
\end{align}
where the final result does not depend on $\xi$ as it multiplies a total derivative term in the third line.
For a scalar with a non-minimal coupling to the curvature we find
\begin{align}
\begin{aligned}
	\frac{1}{d-2}\biggl\langle \frac{\delta \Theta}{\delta\CR}\biggr\rangle &= - \frac{1}{2}\xi\, \langle \phi^2 \rangle \ , \\
		&= - \frac{\xi\,m^{d-2}}{2^{d+1} \pi^{\frac{d}{2}-1} \sin\left( \frac{\pi d}{2}\right) \Gamma(d/2)} \ .
\end{aligned}
\end{align}
We would naively then obtain the (incorrect) universal area term
\begin{align}\label{UniversalArea}
	S|_\text{univ} =  \frac{(1-6\xi)}{3\cdot 2^d \pi^{\frac{d}{2}-2}\sin\left(\frac{\pi d}{2}\right) \Gamma(d/2)} m^{d-2}\CA_\Sigma \ .
\end{align}
As we have argued, the correct result should correspond to the choice $\xi =0$.  Only in this case do we match the 
$d=3$ (\ref{betavalue}) and $d=4$ (\ref{DeltaSd4}) area law results mentioned earlier in the paper. 

We then further compute how the Adler-Zee formula responds to shifting the stress tensor by the equation of motion.  
The two-point function shifts by
\begin{align}
		\int \d^d x\, {\vec x}^{\,2}\, \langle \Theta_{(\kappa)}(x)\Theta_{(\kappa)}(0)\rangle &= \int \d^d x\, {\vec x}^{\,2}\, \langle \Theta(x)\Theta(0)\rangle - \frac{d(d-2)^2 \kappa (\kappa - 2\xi/\xi_c)\,m^{d-2}}{2^{d+1} \pi^{\frac{d}{2}-1}\sin\left(\frac{\pi d}{2}\right) \Gamma(d/2)} \ ,
\end{align}
while the contact term is modified to
\begin{align}
\label{contactR}
	\frac{1}{d-2}\biggl\langle \frac{\delta \Theta_{(\kappa)}}{\delta\CR}\biggr\rangle = \frac{\kappa - 1}{2}\xi\, \langle \phi^2 \rangle \ .
\end{align}
In this prescription, the universal area term is shifted from \eqref{UniversalArea} by a $\kappa$-dependent term:
\begin{align}\label{UniversalAreaOn}
	\delta S|_\text{on-shell, univ} =  \frac{\xi_c\,\kappa(\kappa - \xi/\xi_c)}{2^{d-1} \pi^{\frac{d}{2}-2}\sin\left(\frac{\pi d}{2}\right) \Gamma(d/2)}m^{d-2}\CA_\Sigma \ .
\end{align}
The calculation above duplicates and agrees with one in ref.\ \cite{Casini:2014yca} for the particular choices $\kappa = 0$ and $\kappa = \xi / \xi_c$.  Here, however, we see the general $\kappa$ dependence of the Adler-Zee formula.

We have gone through the exercise of computing the $\kappa$-dependence of the Adler-Zee formula because of an alternate method of computation that relies on a spectral decomposition of the two-point function of the trace of the stress tensor:
\begin{align}
\begin{aligned}
	\langle \Theta(x)\Theta(0)\rangle 	&= \frac{\pi^{\frac{d}{2}}}{(d+1)2^{d-2}\Gamma(d)\Gamma(d/2)}\int_0^\infty \d\mu\, c^{(0)}(\mu)\, (\mu^2)^2\, G(x, \mu) \ ,
\end{aligned}
\end{align}
where $G(x,\mu)$ is the Green's function of a free scalar field on flat space with mass $\mu$ and $c^{(0)}(\mu)$ is the spin zero spectral function \cite{Cappelli:1990yc}.
The area term involving the two-point function $\langle \Theta\, \Theta\rangle$ is nicely written in ref.\ \cite{Casini:2014yca} as
\begin{align}\label{SpectralDecomposision}
	 - \frac{\pi^{\frac{d}{2}+1} }{2^{d-3}(d^2-1)(d-2)\Gamma(d)\Gamma(d/2)}\int_0^\infty \d \mu\, c^{(0)}(\mu) \ .
\end{align}
In this spectral decomposition, for the conformally coupled scalar the equations of motion are typically \cite{Cappelli:1990yc,Ben-Ami:2015zsa} used to reduce the stress tensor to a term proportional to $m^2 \phi^2$.  In our notation, these choices correspond to $\xi = \xi_c$ and $\kappa = 1$.  Moreover, in the spectral computation the contact term proportional to $\delta \Theta / \delta {\mathcal R}$ is ignored.
One might worry, given the $\kappa$-dependence above, that employing the equations of motion and ignoring the contact term will lead to further complications. However, precisely for the choices $\kappa = \xi / \xi_c$ and $\kappa=1$, the $\kappa$-dependent shift \eqref{UniversalAreaOn} vanishes and the contact term (\ref{contactR}) can be ignored, respectively.  The result then matches the (wrong) calculation (\ref{UniversalArea}) with $\xi = \xi_c$.

\subsection{Rosenhaus-Smolkin formula}
Recalling the definition of the modular Hamiltonian $H \equiv - \log \rho$, the entanglement entropy is its expectation value $S = \langle H\rangle$.
The variation of the entropy with respect to the coupling $g_\CO$ of a relevant operator $\CO(x)$ is then given by Rosenhaus-Smolkin in ref.\ \cite{Smolkin:2014hba}:
\begin{align}\label{RSformula}
	g_\CO\,\partial_{g_\CO} S = - g_\CO \int \d^d x\,\sqrt{g}\, \langle \CO(x)\, H \rangle \ .
\end{align}
Note that the expectation value on the right hand side is taken for a state perturbed by the operator.  (Comparing this result to our earlier (\ref{1stS}), one may draw an analogy to the relation between time-independent first order perturbation theory in quantum mechanics and the Feynman-Hellman Theorem.)

As in the preceding sections we add the boundary term in the modular Hamiltonian that contributes to the right hand side of \eqref{RSformula}
\begin{align}
	- g_\CO \int \d^d x\,\sqrt{g}\, \langle \CO(x)\, \delta H \rangle \ .
\end{align}
For a flat entangling surface, the modular Hamiltonian of a non-minimally coupled scalar is improved to be the minimal one by adding the boundary term:
\begin{align}
	H = H_\text{cov} + \delta H \ .
\end{align}
Thus the universal area term should not depend on $\xi$.

We can be more explicit and see how the boundary term $\delta H$ acts to cancel the $\xi$ dependent contribution of the entropy.
For the mass perturbation for a conformally coupled scalar with $g_\CO = m^2$ and $\CO = \phi^2/2$, we find 
\begin{align}
\begin{aligned}
	m^2 \partial_{m^2} \delta S &= - \pi \xi\,m^2 \int \d^d x\, \d^{d-2}y_\perp \langle \phi^2 (x) \phi^2 (y) \rangle \ , \\
		&= \frac{d-2}{2} \frac{\xi}{2^{d-1}\pi^{\frac{d}{2}-2} \sin\left( \frac{\pi d}{2}\right) \Gamma(d/2)}m^{d-2}\CA_\Sigma \ .
\end{aligned}
\end{align}
Integration of both sides yields
\begin{align}
	\delta S = \frac{\xi}{2^{d-1}\pi^{\frac{d}{2}-2} \sin\left( \frac{\pi d}{2}\right) \Gamma(d/2)}m^{d-2}\CA_\Sigma \ .
\end{align}
Adding it to (\ref{UniversalArea}), we obtain the same result as the minimally coupled scalar, i.e.\ (\ref{UniversalArea}) with $\xi=0$.

Interestingly, the boundary term contributes oppositely to the contact term in the Adler-Zee formula without imposing the eom:
\begin{align}
	\delta S = 2\pi\, \xi\,\langle\phi^2 \rangle \CA_\Sigma = \frac{4\pi}{d-2}\biggl\langle \frac{\delta\Theta}{\delta \CR}\biggr\rangle \ ,
\end{align}
Thus our prescription to add the boundary term may be restated as  removing the contact term.

Assuming for the moment the naive definition (\ref{Hdef}) of $H$ that does not include an extra boundary term, we can relate the result (\ref{RSformula}) directly to the Adler-Zee formula (\ref{UniversalArea}).  
Assuming $\CO$ is the only relevant operator in the theory, 
the first step  is to make use of a Ward identity to replace $\langle \CO(x) \, H_{\rm cov} \rangle$ with $\langle \Theta(x)\,  H_{\rm cov}\rangle$:
\be
g_{\CO} \partial_{g_{\CO}} S = \frac{1}{d-\Delta}\left[ \int \d^d x \sqrt{g}\,  \langle \Theta(x)\, H_{\rm cov} \rangle - 4 \pi
\int \d^d x \sqrt{g} \int \d^{d-1} y \, y^1 
 \left \langle \frac{\delta \Theta(x)}{\delta g_{00}(y)} \right \rangle \right]\ ,
\ee
where we have kept a contact term dropped in \cite{Smolkin:2014hba} but continued to ignore anomaly terms as we work in flat space.
The correlator $\langle \Theta(x)\, H_{\rm cov} \rangle$ can then be directly related to the trace-trace correlator in the Adler-Zee formula \cite{Casini:2014yca}.  The contact term can be rewritten as a functional derivative of $\Theta$ with respect to ${\mathcal R}$ on the conical space.
The Rosenhaus-Smolkin formula with the naive $H_{\rm cov}$ (\ref{Hdef}) actually is then equivalent to the Adler-Zee formula with the contact term, yielding the same universal area term \eqref{UniversalArea}  as before \cite{Casini:2014yca}.

Leaving a more detailed discussion of the equivalence between $\langle \Theta\, \Theta \rangle$ and $\langle \Theta\, H_{\rm cov} \rangle$ to ref. \cite{Casini:2014yca}, we discuss instead how to rewrite the contact term.

The variational derivative with respect to $g_{00}$ will produce terms proportional to $\delta (x-y)$ and $\partial_1$ derivatives acting on it.
Terms proportional to $\partial_1 \delta(x-y)$ do not appear.  
The term we care about looks like $\partial_1^2 \delta(x-y)$ and will lead, after a double integration by parts, to a contact term evaluated at the entangling surface $y^1=0$.  Such a term can only come from the variation of the Ricci curvature.  We know
\be
\delta \CR = - \CR^{\mu\nu} \delta g_{\mu\nu} + \nabla^\mu(\nabla^\nu \delta g_{\mu\nu} - g^{\lambda\rho} \nabla_\mu \delta g_{\lambda\rho}) \ .
\ee 
In flat space, we may rewrite (ignoring the $\delta (x-y)$ term)
\begin{align}
	\frac{\delta \Theta(x)}{\delta g_{tt}(y)} = \int \d^d z\, \frac{\delta \Theta(x)}{\delta \CR(z)} \frac{\delta \CR(z)}{\delta g_{tt}(y)} = - \sum_{i\neq 0} \int \d^d z\, \frac{\delta \Theta(x)}{\delta \CR(z)}\, \partial_{y_i}^2 \delta (z-y) \ ,
\end{align}
and find
\begin{align}\label{dTdg}
	\int \d^d x \int_{y_1>0} \d^{d-1} y \, y^1  \left \langle \frac{\delta \Theta(x)}{\delta g_{00}(y)} \right \rangle &= -\int \d^d x \int_{y_1=0} \d^{d-2} y\,\left \langle\frac{\delta \Theta(x)}{\delta \CR(y)} \right \rangle = -\CA_\Sigma\,\left \langle\frac{\delta \Theta}{\delta \CR} \right \rangle \ ,
\end{align}
where the second inequality follows from the assumption that the stress tensor has a curvature dependence through the coupling $\CR\,\CO$ to the relevant operator.
By dimensional analysis, $\langle \delta \Theta/\delta \CR \rangle$ must scale as $g_\CO^{\frac{d-2}{d-\Delta}}$.  Integrating this result with respect to the coupling $g_{\CO}$, we produce a factor of $\frac{d-2}{d-\Delta}$ and match the contact term contribution to (\ref{entropyresult}) on the nose.

Finally, let us comment on the $\delta(x-y)$ term ignored in \eqref{dTdg}.
It is independent of the position from the translational invariance, e.g., $\delta\Theta(x)/\delta g_{00}(y) \sim c\, \delta(x-y)$ with constant $c$, and contributes to \eqref{dTdg} an infrared divergent constant.
Thus such a divergence cannot be universal as it needs to be regularized and renormalized in some way.

\section{Twist Operator of Spherical Entangling Surface}
\label{sec:twist}
The partition function on the replica manifold can be regarded as a correlation function of a codimension-two twist operator $\sigma_\alpha$ located on an entangling surface.
Their characterization is far from our reach in general though. Some aspects have been investigated \cite{Hung:2011nu,Hung:2014npa} for a planar twist operator in CFT.
The tracelessness and conservation of the stress tensor determine the correlation function $\langle T_{\mu\nu}\, \sigma_\alpha\rangle$ to be\footnote{Our convention for the stress tensor in Euclidean signature is $T_{\mu\nu} = \frac{2}{\sqrt{g}} \frac{\delta I_E}{\delta g^{\mu\nu}}$.}
\begin{align}
	\begin{aligned}
		\langle T_{ab}\, \sigma_\alpha\rangle &= \frac{h_\alpha}{2\pi}\frac{\delta_{ab}}{y^d} \ , \qquad \langle T_{ai}\, \sigma_\alpha\rangle = 0 \ , \\
		\langle T_{ij}\, \sigma_\alpha\rangle &= -\frac{h_\alpha}{2\pi}\frac{(d-1)\delta_{ij} - d\, n_i n_j}{y^d}  \ ,
	\end{aligned}
\end{align}
up to a factor $h_\alpha$ which will be called the conformal dimension of the twist operator spanning over the coordinates $x^a$ $(a=3,\cdots, d)$ and sitting on the origin $y^1 = y^2 = 0$.
$n^i$ $(i=1,2)$ is the unit normal vector to the entangling surface.

By the conformal map to $\BS^1 \times \BH^{d-1}$ the conformal dimension takes the form 
\begin{align}\label{hn_def}
	h_\alpha = \frac{2\pi \alpha\, R^d}{d-1} \left( \CE_1 - \CE_\alpha \right) \ ,
\end{align}
with the vacuum energy on $\BH^{d-1}$ at temperature $T = 1/(2\pi R\,\alpha)$
\begin{align}
	\CE_\alpha 
	= - \frac{\text{Tr}\left[ T_{\tau\tau} e^{-\alpha H}\right]}{Z_\alpha}\ .
\end{align}
The minus sign is due to the convention for the Euclidean stress tensor and correspondingly the bulk modular Hamiltonian has minus sign, $H = -2\pi \int_{\BH^{d-1}} \d^{d-1}x\, T_{\tau\tau}$.
As we have already discussed, the modular Hamiltonian $H$ is not simply an integral over the energy density $T_{\tau\tau}$ (\ref{Hdef}) but involves
also a boundary term (\ref{Hfixed}).

Just as in the case of massive perturbation theory considered in the previous section, we have two methods at our disposal for
calculating $h_\alpha$.  The first involves the replica trick and the Plancherel measure on $\BH^{d-1}$.  The second involves calculating correlation 
functions of $H$ with other fields, in this case $T_{\tau\tau}$.  

In fact we will not compute $h_\alpha$ but the first few coefficients in a Taylor series expansion of $h_\alpha$ near $\alpha=1$:
\be\label{hna}
h_{\alpha,a} \equiv \partial_\alpha^a h_\alpha |_{\alpha=1} \ .
\ee
The coefficients are closely related to the derivative of the R{\'e}nyi entropy around $\alpha=1$ \cite{Perlmutter:2013gua}
\begin{align}
	S_{\alpha,a} = \frac{2\pi R}{a+1} E_{\alpha,a} \ ,
\end{align}
where we used the same notation as \eqref{hna} and $E_\alpha$ is the integrated energy 
\begin{align}\label{EErelation}
	E_\alpha \equiv \int_{\BH^{d-1}}\, \CE_\alpha \ .
\end{align}
The homogeneity of $\BH^{d-1}$ allows us to factor out the volume in the right hand side of \eqref{EErelation}.
Then we find a simple relation between $h_{\alpha,a}$ and $S_{\alpha,a}$
\begin{align}\label{hnSnRelation}
	h_{\alpha,a} = - \frac{1}{(d-1)\text{Vol}(\BH^{d-1})} \left[ (a+1)S_{\alpha,a} + a^2(1- \delta_{a,1})\, S_{\alpha,a-1}\right] \ .
\end{align}
For completeness we give the inverse relation 
\begin{align}
	S_{\alpha, a} = - \frac{a!\, (d-1)\text{Vol}(\BH^{d-1})}{a+1}\sum_{k=1}^a (-1)^{a-k}\, \frac{h_{\alpha, k}}{k!} \ .
\end{align}
This is the same relation as the one in \cite{Hung:2014npa} up to $a=2$ and corrects 
the expressions for $a\ge 3$.  Corrected relationships can be found in appendix E of \cite{Bueno:2015qya} along with further discussion of the $h_\alpha$.

With these relations, what we are going to show is equivalent to the extension to general dimensions of ref. \cite{Lee:2014zaa} where a related discrepancy was resolved examining $S_{\alpha,a}$ for $d=3$ and 4 dimensions.

\subsection{Free energy method}
We can evaluate $h_\alpha$ defined by \eqref{hn_def} using the energy density on $\BS^1\times \BH^{d-1}$ which can be obtained by taking a derivative of the free energy \eqref{F_n_scalar} with respect to $1/T = 2\pi R\, \alpha$:
\begin{align}
\begin{aligned}
	h_\alpha &= \frac{\alpha}{(d-1)\Vol(\BH^{d-1})} \left( \partial_\alpha F_\alpha |_{\alpha=1} - \partial_\alpha F_\alpha\right) \ , \\
		&= \frac{\alpha\,\pi}{(d-1)\Vol(\BH^{d-1})}\int_0^\infty \d\lambda\, \mu_s (\lambda)\,\sqrt{\lambda} \left[ \coth (\pi\sqrt{\lambda}) - \coth (\alpha \pi\sqrt{\lambda}) \right] \ .
\end{aligned}
\end{align}
Here $\mu_s(\lambda)$ is the Plancherel measure (\ref{Plancherel}).
It follows immediately that
\begin{align}
	\begin{aligned}
		h_{\alpha,1} &= \frac{1}{2^d\pi^{\frac{d-3}{2}}\Gamma\left( \frac{d+1}{2}\right)}\int_0^\infty \d\lambda\,\frac{\lambda}{\sinh(\pi\sqrt{\lambda})} \bigg| \Gamma\left( \frac{d}{2} -1 + i\sqrt{\lambda}\right) \bigg|^2 \ , \\
		h_{\alpha,2} &= - \frac{1}{2^{d-1}\pi^{\frac{d-3}{2}}\Gamma\left( \frac{d+1}{2}\right)}\int_0^\infty \d\lambda\,\frac{\lambda}{\sinh(\pi\sqrt{\lambda})} \left[ \pi\sqrt{\lambda}\coth (\pi\sqrt{\lambda}) -1)\right] \bigg| \Gamma\left( \frac{d}{2} -1 + i\sqrt{\lambda}\right) \bigg|^2 \ .
	\end{aligned}
\end{align}
These integrals are very similar to one (\ref{Fderiv}) that we did before in computing mass corrections to entanglement entropy.  Again, it is useful to make a change of variables $\lambda = x^2$ and employ the integral formula (\ref{firstBarneslemma}).  Before applying this integral formula, we have to replace the hyperbolic trigonometric functions with $\Gamma$-functions.  The following two relations are useful, the first to evaluate $h_{\alpha,1}$ and the second to evaluate $h_{\alpha,2}$:
  \be
  \frac{\pi x^3}{\sinh (\pi x)} &=&| \Gamma(2+ix)|^2 - |\Gamma(1+ix)|^2\ , \\
  \frac{2 \pi x^3}{\sinh(\pi x)} \left[ \pi x \coth (\pi x) - 1 \right]  &=& \frac{\partial}{\partial n} \left[ \Gamma(3-n-ix) \Gamma(2+n+ix) - \Gamma(3+n+i x) \Gamma(2-n-i x) \right. \nonumber \\
  && \left. ~+ \Gamma(2+n+ix) \Gamma(1-n-ix)  - \Gamma(2-n-ix) \Gamma(1+n+i x) \right]_{n=0} \nonumber \\
  && ~~+ 6 (|\Gamma(2+ix)|^2 - |\Gamma(1+ix)|^2 ) \ ,
  \ee
We obtain then the first two nonzero coefficients of the Taylor series expansion of $h_\alpha$ near $\alpha=1$:
\begin{align}
\label{hn1}
	h_{\alpha,1} &= \frac{\Gamma \left( d/2 \right)^2}{2^d \pi^{\frac{d-3}{2}} (d^2-1)\, \Gamma \left( (d+1)/2 \right)} \ , \\
\label{hn2}
	h_{\alpha,2} &= - \frac{(3d^2 - 2d-4)\Gamma (d/2)^3}{\pi^{\frac{d}{2}-1} (d-1)\,\Gamma (d+3)} \ .
\end{align}

Using a related heat kernel approach, ref.\ \cite{Hung:2014npa} computed $h_{\alpha,1}$ in general and $h_{\alpha,2}$ for even $d$, $4 \leq d \leq 14$.  Our results (\ref{hn1}) and (\ref{hn2}) agree with theirs.  However, ref.\ \cite{Hung:2014npa} noted a discrepancy when they attempted to compute $h_{\alpha,2}$ using the modular Hamiltonian because they did not include the boundary term.  We will now see in detail how the boundary term fixes the discrepancy.

\subsection{Modular Hamiltonian method}

The boundary term in question amounts to shifting the naive modular Hamiltonian (\ref{Hdef}) by
\begin{align}
	\delta H = 2\pi \xi \int_{\partial\BH^{d-1}} \phi^2 \ .
\end{align}
We will repeat the analysis done by \cite{Hung:2014npa} and see if the shift resolves the puzzle.
We are going to calculate the first and second derivatives of the conformal dimensions at $\alpha=1$
\begin{align}
\label{hnmodular}
	\begin{aligned}
		h_{\alpha,1} &= -\frac{2\pi R^d}{d-1} \langle (H_0+\delta H)\, T_{\tau\tau} \rangle \ ,\\
		h_{\alpha,2} &= \frac{2\pi R^d}{d-1}\left[ \langle (H_0+\delta H)(H_0+\delta H)\, T_{\tau\tau} \rangle - 2 \langle (H_0+\delta H) T_{\tau\tau}\rangle\right]\ ,
	\end{aligned}
\end{align}
where $H_0 = H_{\rm cov}$ of the introduction with the opposite sign and the correlators are in Euclidean signature. 
The conformal symmetry imposes the constraint 
\begin{align}
	\langle T_{\mu\nu}(x)\,\CO(y)\rangle = 0 \ ,	
\end{align}
for an (local) operator $\CO$, leading to $\langle \delta H\, T_{\tau\tau}\rangle = 0$.
Thus the first derivative $\partial_\alpha h_\alpha |_{\alpha=1}$ is not affected by the boundary term.  Indeed, ref.\ \cite{Hung:2014npa} compute $h_{\alpha,1}$ using only $H_0$ and the result agrees with our replica calculation (\ref{hn1}).  However, the second derivative deviates from the naive value
\begin{align}
	h_{\alpha,2} &= h_{\alpha,2}^{(0)} + \delta h_{\alpha,2} \ ,
\end{align}
where 
\begin{align}\label{deltah2}
	 \delta h_{\alpha,2} = \frac{2\pi R^d}{d-1 }\left[ 2 \langle \delta H\, H_0\, T_{\tau\tau} \rangle + \langle \delta H\, \delta H\, T_{\tau\tau} \rangle\right]\ .
\end{align}

To evaluate the three-point functions involving the modular Hamiltonian and the counter term, we find it convenient to put the stress tensor at $(\tau, u)=(\pi R, u_1\ll 1)$ and the others at $\tau = 0$ following \cite{Hung:2014npa} where they obtain 
\begin{align}
	h_{\alpha,2}^{(0)} = -\frac{16\pi^{d+1}}{d^2 \,\Gamma (d+3)} \left[ 2(d-2)(3d^2 - 3d - 4)a - 2d(d-1)b - (3d-4)(d+1) c\right] \ .
\end{align}
For a free conformally coupled real scalar, the parameters are given by \cite{Osborn:1993cr}
\begin{align}
	a = \frac{d^3}{8\Vol(\BS^{d-1})^3 (d-1)^3} \ , \qquad b = -\frac{d^4}{8\Vol(\BS^{d-1})^3 (d-1)^3} \ , \qquad c = -\frac{d^2(d-2)^2}{8\Vol(\BS^{d-1})^3 (d-1)^3} \ ,
\end{align}
and we obtain the corresponding conformal dimension (correcting a typo in (3.26) of ref.\ \cite{Hung:2014npa})
\begin{align}\label{hn2raw}
	h_{\alpha,2}^{(0)} = -\frac{\Gamma (d/2)^3 (11d^4 - 33 d^3 + 16d^2 + 28 d - 16)}{4\pi^{\frac{d}{2}-1} \Gamma (d+3) (d-1)^3}  \ ,
\end{align}
which evidently does not agree with the result (\ref{hn2}) from the previous subsection.

We now proceed to evaluate the three-point functions $\langle \delta H\, H_0\, T_{\tau\tau} \rangle$ and $\langle \delta H\,\delta H\, T_{\tau\tau} \rangle$ on $\BS^1 \times \BH^{d-1}$ by using a Weyl transformation from flat space.
The first step is then to calculate the necessary three-point functions in flat space.

\subsubsection*{The three-point functions for a conformally coupled scalar}
 
 It is useful to put the stress tensor in the form ((5.1) of ref.\ \cite{Osborn:1993cr})
 \be
 \label{OsbornTmunu}
 T_{\mu\nu} = (\partial_\mu \phi) (\partial_\nu \phi) + \left[- \xi \partial_\mu \partial_\nu + \left(\xi - \frac{1}{4} \right) \delta_{\mu\nu} \partial^2 \right] \phi^2 \ ,
 \ee
 eliminating the $(\partial \phi)^2$ term.  In general, $(\partial_i \phi) (\partial_j \phi)$ terms are more difficult to deal with because they require point splitting in the analysis.  In this form, in considering $T_{\tau \tau}$, the $(\partial_\tau \phi)^2$ term will not contribute to the three point functions because all of the time coordinates of the insertion points have been set to zero.

As we have done before, we use the normalization of the two-point function
 \be
 \langle \phi(x) \phi(y) \rangle = \frac{1}{(d-2) \Vol(\BS^{d-1})} \frac{1}{|x-y|^{d-2}} \ .
 \ee
 We can then write the three-point functions in the form
  \be
  \label{K1}
 \langle T_{\tau \tau} (x_1) \phi^2(x_2) \phi^2(x_3) \rangle 
 &=&
 \frac{8}{(d-2)^3 \Vol(\BS^{d-1})^3}  \left[- \xi \partial_{\tau_1}^2 + \left(\xi - \frac{1}{4} \right)\partial_1^2  \right] 
 \frac{1}{x_{12}^{d-2} x_{13}^{d-2} x_{23}^{d-2}} \ ,\nonumber \\
 &=&\frac{2}{(d-2) (d-1) \Vol(\BS^{d-1})^3} \frac{1}{x_{12}^d x_{13}^d x_{23}^{d-4}} \ , \\
 \label{K2}
 \langle T_{\tau \tau}(x_1) T_{\tau \tau}(x_2) \phi^2(x_3) \rangle &=&
  \frac{8}{(d-2)^3 \Vol(\BS^{d-1})^3}\left[ \prod_{i=1}^2 \left(- \xi \partial_{\tau_i}^2 + \left(\xi - \frac{1}{4} \right)\partial_i^2  \right) \right]
 \frac{1}{x_{12}^{d-2} x_{13}^{d-2} x_{23}^{d-2}} \ ,\nonumber \\
 &=& \frac{3 d}{2(d-1)^2 \Vol(\BS^{d-1})^3} \frac{1}{x_{12}^{d+2} x_{23}^{d-2} x_{13}^{d-2}} \ .
 \ee
 The factor of 8 comes from contracting the $\phi$ fields in various equivalent ways.

Note we have used the equations of motion to eliminate a $\phi \partial^2 \phi$ term from the stress tensor (\ref{OsbornTmunu}).  In general, this substitution could lead to troublesome contact terms in the three-point functions.  In the case here, we will always move $x_1$ far away from $x_2$ and $x_3$, which eliminates most of the contact term ambiguity.  There remains a possible issue with the limit $x_2 \to x_3$ in the $\langle T_{\tau\tau}(x_1) T_{\tau \tau}(x_2) \phi^2(x_3)\rangle$ correlation function.  However, this contact term is proportional to $\langle T_{\tau \tau}(x_1) \phi^2(x_2) \rangle$, which vanishes in CFT.

\subsubsection*{Evaluation of $\langle \delta H\, \delta H\, T_{\tau\tau} \rangle$}
Starting with the second term, we use a conformal map from $\BS^1\times \BH^{d-1}$ to $\BR^d$ with a spherical entangling ball region $B$ of radius $R$ and reduce it to an integral of the three-point function on a flat space
\begin{align}\label{dHdHT}
\begin{aligned}
	\langle \delta H\, \delta H\, T_{\tau\tau}(\tau=\pi, u_1) \rangle = \left(2 \pi \xi \right)^2\,\left( -\Omega(\vec r_1)\right)^{-d}  &\int_{\partial B}\d^{d-2}\vec r_2\,  \int_{\partial B}\d^{d-2}\vec r_3\,\\
		& \cdot \langle T_{\tau\tau}(t=0, \vec r_1)\, \phi^2(t=0, \vec r_2) \,\phi^2 (t=0, \vec r_3)\rangle \ ,
\end{aligned}
\end{align}
where $\Omega$ is the Weyl factor \eqref{Weylfactor} at $t=0$
\begin{align}
	\Omega (\vec r) = \frac{2R^2}{R^2 - r^2} \ .
\end{align}

We take a limit of $u_1\ll 1$ that corresponds to an $r_1 \gg R$ limit, where the three-point function (\ref{K1}) simplifies
\begin{align}
	\langle T_{\tau\tau}(t=0, \vec r_1)\, \phi^2(t=0, \vec r_2) \,\phi^2 (t=0, \vec r_3)\rangle ~\underset{r_1 \to \infty}{\longrightarrow} ~\frac{K_1}{r_1^{2d} |\vec r_2 - \vec r_3|^{d-4}} \ ,
\end{align}
where 
\begin{align}
	K_1 = \frac{2}{(d-1)(d-2)\Vol(\BS^{d-1})^3} \ .
\end{align}

The spherical symmetry allows us to fix one of the positions $\vec r_2$ to the north pole on a sphere and factor out the volume, leading to
\begin{align}
\begin{aligned}
	\langle \delta H\, \delta H\, T_{\tau\tau}(\tau=\pi, u_1) \rangle &= \frac{\pi^2 \xi^2 K_1 \Vol(\BS^{d-2}) \Vol(\BS^{d-3})}{2^{d-2}R^{d}}\, \int_0^\pi \d\theta\frac{\sin^{d-3}\theta}{\left( 2\sin \frac{\theta}{2} \right)^{d-4}} \ ,\\
		&= \frac{8\pi^{d} \xi^2 K_1}{R^{d}\Gamma(d-1)} \ .
\end{aligned}
\end{align}

\subsubsection*{Evaluation of $\langle \delta H\, H_0\, T_{\tau\tau} \rangle$}
We proceed to evaluate the first term in a similar way
\begin{align}\label{dHHT}
\begin{aligned}
	\langle \delta H\, H_0\, T_{\tau\tau}(\tau=\pi, u_1) \rangle =-4 \pi^2 R\, \xi \,\left( -\Omega(\vec r_1)\right)^{-d}  &\int_{B}\d^{d-1}\vec r_2\, \Omega^{-1} (\vec r_2) \int_{\partial B}\d^{d-2}\vec r_3\,\\
		& \cdot \langle T_{\tau\tau}(t=0, \vec r_1)\, T_{\tau\tau}(t=0, \vec r_2) \,\phi^2 (t=0, \vec r_3)\rangle \ .
\end{aligned}
\end{align}
In the $r_1\gg R$ limit, the three-point function (\ref{K2}) takes the form
\begin{align}
	\langle T_{\tau\tau}(t=0, \vec r_1)\, T_{\tau\tau}(t=0, \vec r_2) \,\phi^2 (t=0, \vec r_3)\rangle ~\underset{r_1 \to \infty}{\longrightarrow} ~\frac{K_2}{r_1^{2d} |\vec r_2 - \vec r_3|^{d-2}} \ ,
\end{align}
where
\begin{align}
	K_2 = \frac{3d}{2(d-1)^2 \Vol(\BS^{d-1})^3} \ .
\end{align}

Plugging into \eqref{dHHT}, we find
\begin{align}
\begin{aligned}
	\langle \delta H\, H_0\, T_{\tau\tau}(\tau=\pi, u_1\ll 1) \rangle &= -\frac{\pi^2 \xi\, K_2}{2^{d-1} R^{d}} \int \d^{d-2}\Omega_2\int \d^{d-2}\Omega_3 \int_0^1 \d x_2\, \frac{x_2^{d-2}(1-x_2^2)}{|\vec x_2 - \vec x_3|^{d-2}} \ , \\
		&= -\frac{8\pi^{d} \xi\, K_2}{3 R^{d}d\,\Gamma (d-1)} \ ,
\end{aligned}
\end{align}
where we utilized the formulae in appendix A of ref.\ \cite{Hung:2014npa}.

\subsubsection*{Final result $\delta h_{\alpha,2}$}
Assembling the pieces with \eqref{deltah2}, we find 
\begin{align}
\begin{aligned}
	\delta h_{\alpha,2} &= \frac{16\pi^{d+1} \xi}{\Gamma(d)}\left( \xi K_1 - \frac{2K_2}{3d} \right) \ , \\
	&= - \frac{(d-2)\,\Gamma (d/2)^3}{4\pi^{\frac{d}{2}-1}(d-1)^3\Gamma (d)} \ .
\end{aligned}
\end{align}
Adding the correction to the naive result \eqref{hn2raw}, we find perfect agreement with the result (\ref{hn2}) from the free energy computation.

\begin{figure}
\begin{center}
	\begin{subfigure}[b]{0.45\textwidth}
		\includegraphics[width=2.9in]{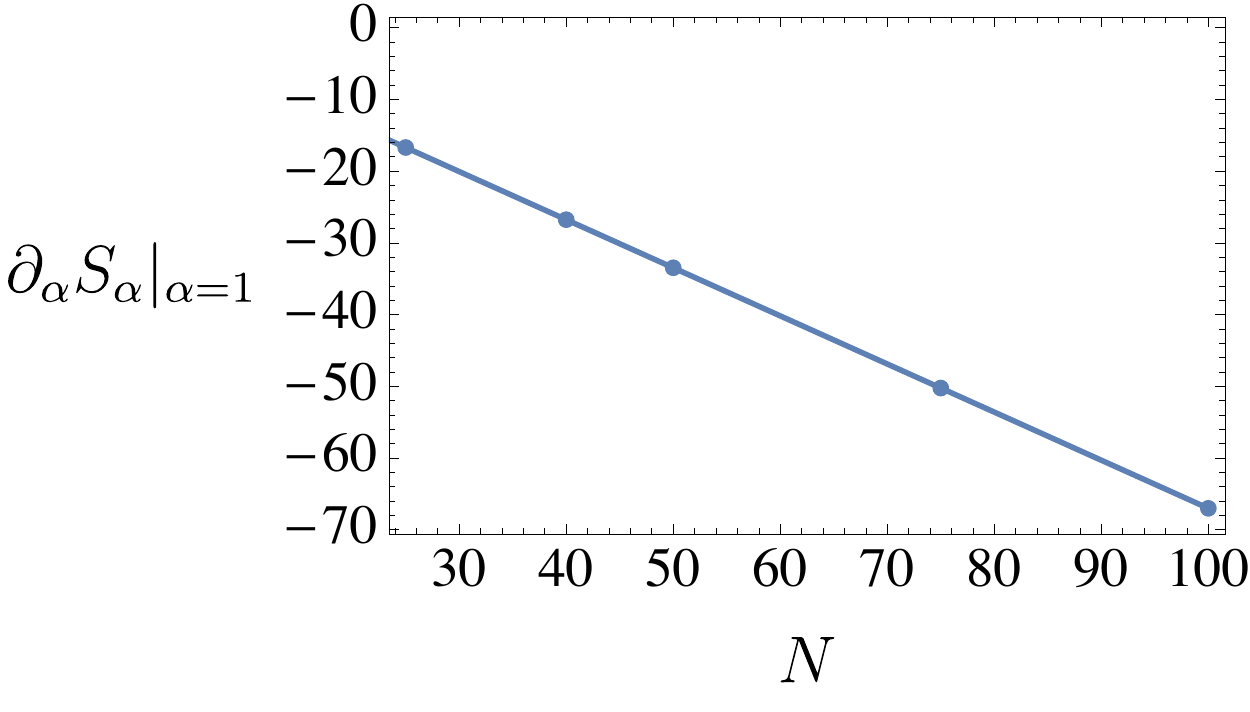}
		\caption{$h_{\alpha,1}$}
	\end{subfigure}
	\hspace*{1cm}
	\begin{subfigure}[b]{0.45\textwidth}
		\includegraphics[width=2.9in]{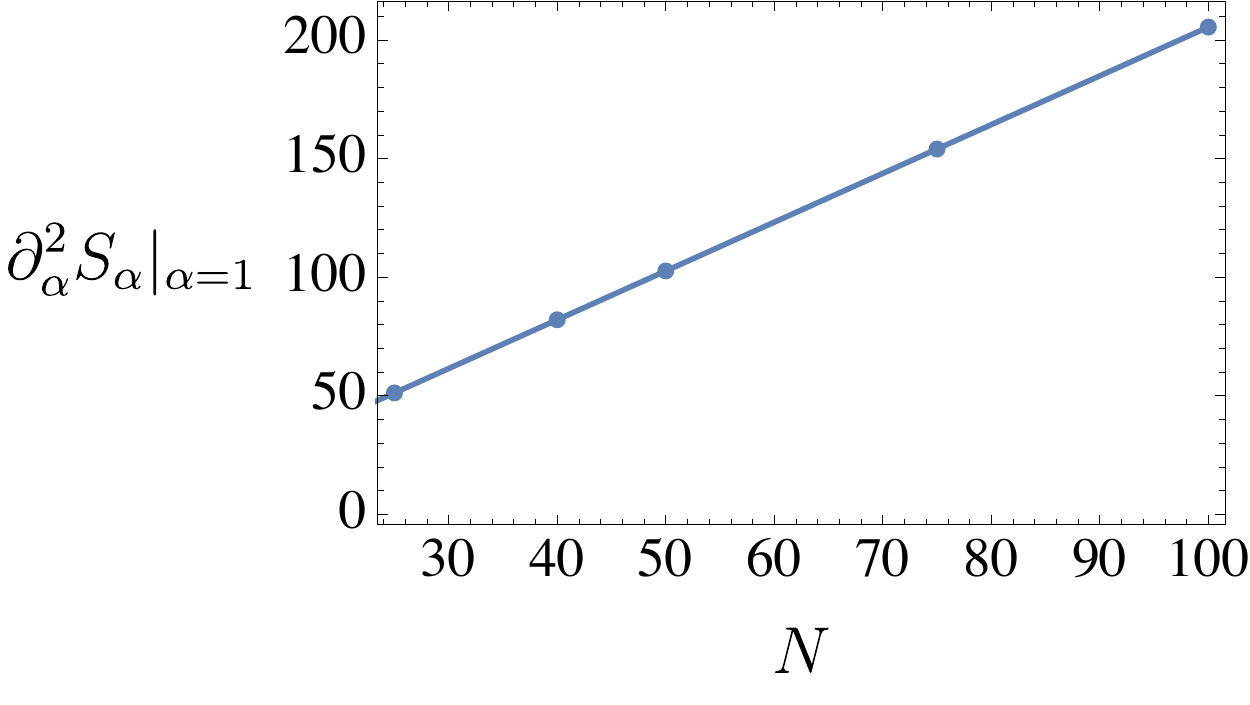}
		\caption{$h_{\alpha,2}$}
	\end{subfigure}
\end{center}
\caption{
Linear fits of $\partial_\alpha S_\alpha |_{\alpha=1}$ (left) and $\partial_\alpha^2 S_\alpha |_{\alpha=1}$ (right) at different lattice spacings in order to compare with $h_{\alpha,1}$ and $h_{\alpha,2}$.  $N$ is the number of lattice points in the entangling region.  For all data points, the entanglement across the equator of $\BS^2$ was computed.  Fitting only the $N=75$ and $N=100$ data points, the fit values are $(-0.671) N  + 0.077$ (left) and $(2.057) N - 0.220$ (right).  Analytically, the intercepts should be at $\pi^2 / 128 = 0.0771063$ and $\pi^2/45 = 0.219325$.  
 \label{fig:twist}
}
\end{figure}

Our numerics confirm that the twist operator dimensions (\ref{hn1}) and (\ref{hn2}) are correct.  See figure \ref{fig:twist}.

\section{Discussion}
\label{sec:discussion}

There are a number of entanglement computations in conformal field theory where one naively might hope to take advantage of the universality of two- and three-point functions involving the stress tensor.  We considered two important examples in this paper.  One was perturbation by a relevant operator ${\mathcal O}$ (in our case a mass term).  The second was a computation of the R\'enyi entropies in the limit $\alpha \to 1$.  
In both cases, we found that the boundary term for $H$, in other words the fact that $H$ was not simply a spatial integral over $T_{00}$, spoiled this expectation.

In the case of perturbation by a relevant operator,  the first order corrections are controlled by $\langle H\, {\mathcal O} \rangle$.
Because $\langle T_{\mu\nu}\, {\mathcal O} \rangle = 0$ in flat space, one might hope to claim in general that first order corrections due to relevant operators vanish, that entanglement entropy is in some sense stationary at a UV conformal fixed point \cite{Rosenhaus:2014woa}.
To make this more precise in 3$d$, we can introduce the quantity $\CF = (L\partial_L -1)S(L)$ for entanglement across a circle of radius $L$ \cite{Liu:2012eea}. 
At a conformal fixed point, $\CF$ corresponds to the subleading constant term in a large $L$ expansion of the entanglement entropy.
Ref.\ \cite{Casini:2012ei}, assuming sub-additivity of the entanglement entropy, demonstrated that $\CF$ decreases monotonically as $L$ increases, like the Zamolodchikov $c$-function for 2$d$-CFTs \cite{Zamolodchikov:1986gt}.  However, refs.\ \cite{Klebanov:2012va,Nishioka:2014kpa} later pointed out that, unlike the $c$-function, $\CF$ was not stationary at the UV fixed point for a conformally coupled scalar perturbed by a mass term.  Here, we see why.  The boundary term in the modular Hamiltonian means that even though $\langle T_{\mu\nu}\, {\mathcal O} \rangle$ is zero, $\langle H\, {\mathcal O} \rangle$ is not.

In the case of R\'enyi entropies, we saw that the first couple of derivatives of $S_\alpha$ in a Taylor series expansion near $\alpha = 1$ were controlled by the two- and three-point functions $\langle H\, T_{00} \rangle$ and $\langle H\, H\, T_{00} \rangle$.  Again, since the two- and three-point functions of the stress tensor are unique up to a couple of undetermined constants, one could hope that $\partial_\alpha S_\alpha|_{\alpha = 1}$ and $\partial_\alpha^2 S_\alpha|_{\alpha = 2}$  have a universal structure. Since the boundary term does not contribute to   $\langle H \,T_{00} \rangle$ for a conformally coupled scalar, this expectation was born out for $\partial_\alpha S_\alpha|_{\alpha = 1}$.  On the other hand, for the next derivative, $\partial_\alpha^2 S_\alpha|_{\alpha = 2}$, it was crucial to include the boundary term.

A third example where the boundary term spoils universality is in thermal corrections to entanglement entropy.  We did not discuss that case in detail here because the issue was already successfully resolved by one of the authors \cite{Herzog:2014fra}.   The correction is governed by $\langle H\, {\mathcal O}\, {\mathcal O} \rangle$ where ${\mathcal O}$ is a primary operator that creates the first excited state, which in turn is naively given by a spatial integral over $\langle T_{00}\, {\mathcal O}\, {\mathcal O} \rangle$, which has a universal form.  However, in this third case as well, the boundary term changes the result.  

A burning question is to what extent these boundary terms could be an issue beyond the case of a conformally coupled scalar.  
We saw already in section \ref{ss:universal} that the same boundary term continues to plays a role for a massive scalar with an arbitrary non-minimal coupling.  Ref.\ \cite{Casini:2014yca} takes the point of view that these boundary terms are very special, that they only can occur when there is an operator in the theory with dimension exactly $\Delta = d-2$ that can be added as a contact term on the entangling surface without spoiling Weyl invariance.  Any additional interaction in the scalar case, the argument goes, would shift the dimension of $\phi^2$ away from $d-2$ and remove the contact term.

While we are sympathetic to this argument, we are troubled that two of the simplest  conformal field theories have issues with boundary terms.  The second example is a U$(1)$ Maxwell field in $3+1$ dimensions.  If we perform BRST quantization and choose Feynman gauge, the BRST action can be written (on a manifold $\CM$ without boundary) as
\be
I = -  \int_\CM \d^4 x\, \sqrt{g}\,\left[ \frac{1}{2} A_\mu (g^{\mu\nu} \Box - \CR^{\mu\nu})A_\nu  +c\, \Box\, b\right]  \ ,
\ee
where $b$ and $c$ are the usual anti-commuting ghost fields.  If we consider a conical space $\CM = C_\alpha \times {\mathbb R}^2$, then the Ricci tensor has a distributional support at the conical singularity,  
$\CR_{ij} = 2 \pi (1-\alpha) \delta_{ij} \delta^2(x)$ where $i, j = 1,2$ index the cone $C_\alpha$.  Analogous to the scalar case we considered in section \ref{sec:bryterm}, one can find eigenfunctions of the vector Laplacian on the cone that involve Bessel functions.  To keep the $J_0(\lambda r)$ eigenfunctions, one has to eliminate the delta function at the origin.  The simplest way to do so is to add a codimension-two boundary term to the action
\be
\delta I = - \pi (1-\alpha) \int_{r=0} \d^2 x\, \delta^{ij} A_i A_j \ .
\ee
Such a boundary term has been discussed before, for example in ref.\ \cite{Kabat:1995eq,Donnelly:2012st}.
Like the boundary term for the scalar, this boundary term for the gauge field is also Weyl invariant.  It is not gauge invariant, but we already broke gauge invariance by choosing Feynman gauge.  Whether the boundary term can be made BRST invariant in a nontrivial way is a subtle question.  In fact boundary conditions in general are a subtle question in computing entanglement entropy.  The usual boundary conditions chosen in quantizing a gauge field on a manifold with boundary do not appear to be compatible with an entanglement computation \cite{Donnelly:2014fua,Donnelly:2015hxa,Huang:2014pfa,Soni:2016ogt}.
More generally, it seems a boundary term $|A_{n-1}|^2$ could appear for free field theories in even dimension $d=2n$ involving an anti-symmetric $n$-form $F_{n} = \d A_{n-1}$ with Lagrangian density $F \wedge \star F$.   

In the CFT context, modular Hamiltonians can be modified by a boundary counter term when there exists an operator $\CO_{d-2}$ of dimension $d-2$ that can have a coupling $\CR\, \CO_{d-2}$.  Moving away from CFT, one no longer needs to ensure that the boundary term is Weyl invariant.  Conceivably any relevant perturbation of the form $\CR \, \CO$ might necessitate a boundary counter term.  In the case of half-space entanglement, the consequences of such a boundary term can be dealt with simply -- the boundary term simply removes the added $\CR \, \CO$ bulk term, leaving one with the original theory.

\acknowledgments
We would like to thank D.\,Fursaev, K.\,Jensen, M.\,Mezei, Y.\,Nakaguchi, S.\,Pufu, J.\,Virrueta, and I.\,Yaakov for discussion.
We would also like to thank the Yukawa Institute for Theoretical Physics at Kyoto University for hospitality, where this work was initiated during the YITP long term workshop on ``Quantum Information in String Theory and Many-body Systems".
C.\,P.\,H. was supported in part by NSF Grants No.\,PHY13-16617 and PHY16-20628.
T.\,N. was supported in part by JSPS Grant-in-Aid for Young Scientists (B) No.\,15K17628.

\appendix

\section{Numerics}
\label{sec:numerics}

We use the same numerical algorithm employed in ref.\ \cite{Herzog:2014fra}.  
The Hamiltonian for a massive scalar on an ${\mathbb S}^{d-1}$ of radius $R$ can be written as a sum over spherical harmonics on an ${\mathbb S}^{d-2}$.  
\be
H = \sum_{\vec l} H_{\vec l} \ ,
\ee
where we index the spherical harmonics by a vector $\vec l$ such that
$|l_1 | \leq l_2 \leq \cdots \leq l_{d-2} \equiv \ell$.  
The spectrum of $H_{\vec l}$ depends only on the largest $l_{d-2} = \ell$:
\be
H_{\vec l} = \frac{1}{2 R^2} \int_{-1}^1 \d u\,\left[ R^2 \pi_{\vec l}^2 - \phi_{\vec l} \,{\mathcal D}\phi_{\vec l} \right]  \ ,
\ee
where
\be
{\mathcal D} \phi_{\vec l} = \partial_u \left((1-u^2) \partial_u \phi_{\vec l}\right) - \frac{\left( \ell + \frac{d-3}{2} \right)^2}{1-u^2} \phi_{\vec l}\, + \left( m^2 - \frac{1}{4} \right) \phi_{\vec l} \ .
\ee
The scalar field $\phi$ has been decomposed into spherical harmonics $\phi_{\vec l}$ as well along with the canonically conjugate 
$\pi_{\vec l}$, such that
\be
[ \phi_{\vec l}\, (u) , \pi_{\vec l'}(u')] = i \delta_{\vec l, \vec l'}\, \delta(u-u') \ .
\ee
We have kept the polar angle on the ${\mathbb S}^{d-1}$ explicit, $ \theta = \cos^{-1} u$, in order to be able to do entanglement computations.
Because ${\mathbb S}^{d-1}$ has no boundary, there is no issue with possible codimension one or two boundary terms in the definition of the full Hamiltonian on this space.  

The R\'enyi entropies can then be constructed from two-point functions of $\phi_{\vec l}$ and $\pi_{\vec l}$ restricted to a cap on the sphere.  The essential observation is that these two-point functions should be the same, whether they are computed with the full or the reduced density matrix.  In this relatively simple case of a scalar field, it is then possible to reconstruct the eigenvalues of the reduced density matrix from the restricted two-point functions.  In technical detail, we define the matrix
\be
C_{\ell}(u_1, u_2)^2 \equiv \int_u^{1}  \d u\,\langle \phi_{\vec l}\, (u_1) \phi_{\vec l}\,(u) \rangle \langle \pi_{\vec l}(u) \pi_{\vec l}(u_2) \rangle\ ,
\ee
where the range of $C_\ell$ is restricted such that $u \leq u_i \leq 1$, $i=1,2$.  The R\'enyi entropy contribution from $H_{\vec l}$ to $S_{\alpha}$ is then
\be
\label{REcont}
R_\alpha(\ell) = \frac{1}{\alpha-1} \tr \left[ \left( C_\ell + \frac{1}{2} \right)^\alpha - \left( C_\ell - \frac{1}{2} \right)^\alpha \right] \ .
\ee
In practice, we do not use this formula itself, but instead use the first few coefficients in a Taylor series near $\alpha = 1$:
\be
R_\alpha(\ell)&=& \tr \left[ \left( C_\ell + \frac{1}{2} \right) \log \left( C_\ell + \frac{1}{2} \right) 
- \left( C_\ell - \frac{1}{2} \right) \log \left( C_\ell - \frac{1}{2} \right)  \right.  \nonumber \\
&&\left.  + \frac{\alpha-1}{8} (1-4 C_\ell^2 ) \log^2 \frac{C_\ell-\frac{1}{2}}{C_\ell + \frac{1}{2}} 
 + \frac{(\alpha-1)^2}{12}C_\ell(1 - 4 C_\ell^2) \log^3 \frac{C_\ell-\frac{1}{2}}{C_\ell + \frac{1}{2}} \right] \ .
\ee
The zeroth order term is the entanglement entropy contribution, which we use in the numerical check of the mass computations in sections \ref{sec:mass} and \ref{sec:largemass}.  The first and second order terms are useful in computing $\partial_\alpha S_\alpha$ and $\partial_\alpha^2 S_\alpha$ near $\alpha = 1$ in the numerical check of the twist operator dimensions in section \ref{sec:twist}.

The total R\'enyi entropy is given by the infinite sum
\be
S_\alpha = R_\alpha(0) + \sum_{\ell=1}^\infty \dim(\ell) R_\alpha(\ell) \ ,
\ee
where $\dim(\ell)$ is the number of spherical harmonics $\vec l$ with $l_{d-2} = \ell$, e.g.\  2 in $d=3$, $2\ell+1$ in $d=4$, etc.
This infinite sum has to be treated with care.  
In the numerics, we perform the sum explicitly up to some $\ell_{\rm max}$ where
$\dim(\ell) R_\alpha(\ell)$ is of order $10^{-7}$.  Then we fit a dozen values of $S_\ell$ with $\ell > \ell_{\rm max}$ to a power law $a \ell^b$, and compute a correction to the finite sum by integrating the power law out to $\ell = \infty$.

To discretize $u$, we choose a grid with lattice points at $u_j = -1 + (j-\frac{1}{2})\epsilon$, $j = 1, \ldots, N$, and $\epsilon = 2/N$.  The operator ${\mathcal D}$ is discretized using
\be
\partial_u((1-u^2) \partial_u f) &\approx& \frac{1}{\epsilon^2} \left[ f_{j-1} \left( 1 - \left( \frac{u_{j-1} + u_j}{2} \right)^2 \right) + f_{j+1} \left( 1 - \left( \frac{u_j + u_{j+1}}{2} \right)^2 \right) \right. \nonumber  \\
&& \quad+ \left. f_j \left( - 2 + \left( \frac{u_{j-1} + u_j}{2}\right)^2 + \left( \frac{u_j + u_{j+1}}{2} \right)^2 \right) \right] \ ,
\ee
valid at second order in $\epsilon$.  With this choice, the contributions from the ghost points $u_0$ and $u_{N+1}$ vanish.  

\section{Further Methods for Computing Mass Corrections}
\label{sec:further}

 \subsection{Method of images (Cardy's method)}
 
 The quantity
 \be
 \frac{1}{\Vol(\BS^1 \times \BH^{d-1})} \lim_{\alpha\to 1} \partial_\alpha \partial_{m^2} (F_\alpha - \alpha\, F_1)|_{m^2 =0} \ ,
 \ee
 can alternately be interpreted as a two-point function on the hyperbolic space near $\alpha=1$:
 \be
\frac{1}{2} \lim_{\alpha \to 1} \partial_\alpha \langle {:} \phi(x)^2 {:} \rangle_{\BS^1 \times \BH^{d-1}} \ .
 \ee
 
We have a closely related quantity from \cite{Herzog:2014tfa}, namely the two-point function $G_{\alpha,d} (2 \theta) = \langle \phi(y) \phi(y') \rangle$ on the conical space $C_\alpha \times {\mathbb R}^{d-2}$ for particular insertion points.  The insertion points $y$ and $y'$ are at the origin of ${\mathbb R}^{d-2}$, at a distance $r/2$ from the origin of $C_\alpha$, and at an angular separation of $2 \theta$.  We will add a factor of $\Vol(\BS^{d-1}) (d-2)$ to change the normalization conventions of the two-point function to the one considered here.  Near $\alpha = 1$, the Green's function has the form
 \be
 G_{(\alpha,d)}(\theta) = G_{(1,d)}(\theta) + (\alpha-1)\, \delta G_{(d)}(\theta) + O(\alpha-1)^2 \ ,
 \ee 
 where
 \be
 G_{(1,d)}(\theta) &=& \frac{1}{(d-2) \Vol(\BS^{d-1})} \frac{1}{\left(2r \sin \frac{\theta}{2}\right)^{d-2}} \ ,\\
 \delta G_{(3)}(\theta) &=& -\frac{1}{\Vol(\BS^2)} \frac{\pi}{8r} \frac{1}{\cos^2 \frac{\theta}{4}} \ ,\\
 \delta G_{(4)}(\theta) &=& - G_{(1,4)}(\theta) \left(-2 + \theta \cot \frac{\theta}{2} \right) \ .
 \ee
 To obtain $\delta G$ for larger $d$, we may use the recursion relation
 \be
 -\frac{\Vol(\BS^{d+1})}{\Vol(\BS^{d-1})} (2r \sin \theta)^2  \delta G_{(d+2)}(2 \theta) +\delta G_{(d)}(2 \theta) = - \pi  \frac{\Vol(\BS^{d-2})}{\Vol(\BS^{d-1})^2} \frac{1}{(d-1)(d-2)} \frac{1}{r^{d-2}}\ .
 \ee
 It follows from this recursion relation that
 \be
 \label{interGd0}
 \delta G_{(d)}(0) = - \pi  \frac{\Vol(\BS^{d-2})}{\Vol(\BS^{d-1})^2} \frac{1}{(d-1)(d-2)} \frac{1}{(2r)^{d-2}} \ .
 \ee
 The corresponding two-point function on $\BS^1 \times \BH^{d-1}$ should be independent of position.  Therefore, the conformal transformation should act to cancel the $1/r$ dependence.  Indeed it does.  We find from this Green's function that,
  \be
\lim_{\alpha \to 1}  \partial_\alpha \langle {:} \phi^2 {:} \rangle_{\BS^1 \times \BH^{d-1}} &=&  -\pi  \frac{\Vol(\BS^{d-2})}{\Vol(\BS^{d-1})^2} \frac{2^{2-d}}{(d-1)(d-2)}\ , \\
&=& - \frac{2^{-d} \pi^{(1-d)/2} \Gamma(d/2)^2}{(d-2) \Gamma((d+1)/2)} \ .
 \ee
 This result agrees with the earlier free energy computation of this two-point function.   
 
\subsection{Contact term from the path integral}
 
There is yet one more way of deriving $\lim_{\alpha \to 1}  \partial_\alpha \langle {:} \phi^2 {:} \rangle_{\BS^1 \times \BH^{d-1}}$.
Without the additional boundary term (\ref{Ibryterm}), the quantity $\langle {:} \phi^2 {:} \rangle_{\BS^1 \times \BH^{d-1}}$ should vanish at $O(1-\alpha)$ for the same reasons that $\langle T^{\mu\nu} {:} \phi^2 {:} \rangle$ vanishes on the plane.  The $O(\alpha-1)$ correction to $\langle {:} \phi^2 {:} \rangle_{\BS^1 \times \BH^{d-1}}$ will thus come purely from this boundary term.  
\begin{align}
	I [\CM_\alpha] =  I_{\rm conf} [\CM] + 2\pi \xi\, (\alpha -1)\int_{r=0}\d^{d-2} y\, \phi^2 (y) + O\left( (\alpha - 1)^2 \right) \ ,
\end{align}
working with the Euclidean action.
Together with the path integral
\begin{align}
	\begin{aligned}
		 G_{(\alpha,d)}(\theta) = \langle {:} \phi^2 (x) {:} \rangle_{\CM_\alpha} &= \int \CD \phi \, \phi^2 (x)\, e^{- I[\CM_\alpha]} \bigg / \left( \int \CD \phi \, e^{- I[\CM_\alpha]} \right)\ , 
	\end{aligned}
\end{align}
 we would get then that
 \begin{align}
 \begin{aligned}
 \delta G_{(d)}(0) &= -2 \pi \xi \int_{\mathbb R^{d-2}} \d^{d-2} y\, \langle {:} \phi^2 (x) {:} {:} \phi^2 (y) {:} \rangle \ ,\\
 &= -\frac{4 \pi \xi \Vol(\BS^{d-3})}{(d-2)^2 \Vol(\BS^{d-1})^2} \int_0^\infty \d y\,\frac{y^{d-3}}{(y^2 + r^2)^{d-2}} \ , \\
 &= -\frac{(d-2) \pi^{(1-d)/2} 2^{-2-d} \Gamma(d/2 - 1)^2}{\Gamma((1+d)/2)} \frac{1}{r^{d-2}} \ ,
\end{aligned}
\end{align}
in agreement with (\ref{interGd0}).

\bibliographystyle{JHEP}
\bibliography{scalarfieldredux}

\providecommand{\href}[2]{#2}\begingroup\raggedright\begin{thebibliography}{10}

\bibitem{Bombelli:1986rw}
L.~Bombelli, R.~K. Koul, J.~Lee, and R.~D. Sorkin, {\it {A Quantum Source of
  Entropy for Black Holes}},  {\em Phys. Rev.} {\bf D34} (1986) 373--383.

\bibitem{Srednicki:1993im}
M.~Srednicki, {\it {Entropy and Area}},  {\em Phys.Rev.Lett.} {\bf 71} (1993)
  666--669, [\href{http://arxiv.org/abs/hep-th/9303048}{{\tt hep-th/9303048}}].

\bibitem{Casini:2006es}
H.~Casini and M.~Huerta, {\it {A C-Theorem for the Entanglement Entropy}},
  {\em J.Phys.} {\bf A40} (2007) 7031--7036,
  [\href{http://arxiv.org/abs/cond-mat/0610375}{{\tt cond-mat/0610375}}].

\bibitem{Casini:2012ei}
H.~Casini and M.~Huerta, {\it {On the RG Running of the Entanglement Entropy of
  a Circle}},  {\em Phys. Rev.} {\bf D85} (2012) 125016,
  [\href{http://arxiv.org/abs/1202.5650}{{\tt arXiv:1202.5650}}].

\bibitem{Solodukhin:2008dh}
S.~N. Solodukhin, {\it {Entanglement Entropy, Conformal Invariance and
  Extrinsic Geometry}},  {\em Phys. Lett.} {\bf B665} (2008) 305--309,
  [\href{http://arxiv.org/abs/0802.3117}{{\tt arXiv:0802.3117}}].

\bibitem{Komargodski:2011vj}
Z.~Komargodski and A.~Schwimmer, {\it {On Renormalization Group Flows in Four
  Dimensions}},  {\em JHEP} {\bf 1112} (2011) 099,
  [\href{http://arxiv.org/abs/1107.3987}{{\tt arXiv:1107.3987}}].

\bibitem{Osborne:2002zz}
T.~J. Osborne and M.~A. Nielsen, {\it {Entanglement in a Simple Quantum Phase
  Transition}},  {\em Phys. Rev.} {\bf A66} (2002) 032110.

\bibitem{Vidal:2002rm}
G.~Vidal, J.~Latorre, E.~Rico, and A.~Kitaev, {\it {Entanglement in Quantum
  Critical Phenomena}},  {\em Phys.Rev.Lett.} {\bf 90} (2003) 227902,
  [\href{http://arxiv.org/abs/quant-ph/0211074}{{\tt quant-ph/0211074}}].

\bibitem{Kitaev:2005dm}
A.~Kitaev and J.~Preskill, {\it {Topological Entanglement Entropy}},  {\em
  Phys.Rev.Lett.} {\bf 96} (2006) 110404,
  [\href{http://arxiv.org/abs/hep-th/0510092}{{\tt hep-th/0510092}}].

\bibitem{Levin:2006zz}
M.~Levin and X.-G. Wen, {\it {Detecting Topological Order in a Ground State
  Wave Function}},  {\em Phys.Rev.Lett.} {\bf 96} (2006) 110405.

\bibitem{Bisognano:1976za}
J.~J. Bisognano and E.~H. Wichmann, {\it {On the Duality Condition for Quantum
  Fields}},  {\em J. Math. Phys.} {\bf 17} (1976) 303--321.

\bibitem{Bisognano:1975ih}
J.~J. Bisognano and E.~H. Wichmann, {\it {On the Duality Condition for a
  Hermitian Scalar Field}},  {\em J. Math. Phys.} {\bf 16} (1975) 985--1007.

\bibitem{Casini:2011kv}
H.~Casini, M.~Huerta, and R.~C. Myers, {\it {Towards a Derivation of
  Holographic Entanglement Entropy}},  {\em JHEP} {\bf 05} (2011) 036,
  [\href{http://arxiv.org/abs/1102.0440}{{\tt arXiv:1102.0440}}].

\bibitem{Wong:2013gua}
G.~Wong, I.~Klich, L.~A. Pando~Zayas, and D.~Vaman, {\it {Entanglement
  Temperature and Entanglement Entropy of Excited States}},  {\em JHEP} {\bf
  1312} (2013) 020, [\href{http://arxiv.org/abs/1305.3291}{{\tt
  arXiv:1305.3291}}].

\bibitem{Barnich:2001jy}
G.~Barnich and F.~Brandt, {\it {Covariant Theory of Asymptotic Symmetries,
  Conservation Laws and Central Charges}},  {\em Nucl. Phys.} {\bf B633} (2002)
  3--82, [\href{http://arxiv.org/abs/hep-th/0111246}{{\tt hep-th/0111246}}].

\bibitem{Fursaev:1998hr}
D.~V. Fursaev, {\it {Energy, Hamiltonian, Noether Charge, and Black Holes}},
  {\em Phys. Rev.} {\bf D59} (1999) 064020,
  [\href{http://arxiv.org/abs/hep-th/9809049}{{\tt hep-th/9809049}}].

\bibitem{Frolov:1997up}
V.~P. Frolov and D.~V. Fursaev, {\it {Mechanism of Generation of Black Hole
  Entropy in Sakharov's Induced Gravity}},  {\em Phys. Rev.} {\bf D56} (1997)
  2212--2225, [\href{http://arxiv.org/abs/hep-th/9703178}{{\tt
  hep-th/9703178}}].

\bibitem{Herzog:2014fra}
C.~P. Herzog, {\it {Universal Thermal Corrections to Entanglement Entropy for
  Conformal Field Theories on Spheres}},  {\em JHEP} {\bf 10} (2014) 28,
  [\href{http://arxiv.org/abs/1407.1358}{{\tt arXiv:1407.1358}}].

\bibitem{Lee:2014zaa}
J.~Lee, A.~Lewkowycz, E.~Perlmutter, and B.~R. Safdi, {\it {R\'enyi Entropy,
  Stationarity, and Entanglement of the Conformal Scalar}},  {\em JHEP} {\bf
  03} (2015) 075, [\href{http://arxiv.org/abs/1407.7816}{{\tt
  arXiv:1407.7816}}].

\bibitem{Casini:2014yca}
H.~Casini, F.~D. Mazzitelli, and E.~Test\'e, {\it {Area Terms in Entanglement
  Entropy}},  {\em Phys. Rev.} {\bf D91} (2015), no.~10 104035,
  [\href{http://arxiv.org/abs/1412.6522}{{\tt arXiv:1412.6522}}].

\bibitem{Metlitski:2009iyg}
M.~A. Metlitski, C.~A. Fuertes, and S.~Sachdev, {\it {Entanglement Entropy in
  the $O(N)$ Model}},  {\em Phys. Rev.} {\bf B80} (2009), no.~11 115122,
  [\href{http://arxiv.org/abs/0904.4477}{{\tt arXiv:0904.4477}}].

\bibitem{Hung:2014npa}
L.-Y. Hung, R.~C. Myers, and M.~Smolkin, {\it {Twist Operators in Higher
  Dimensions}},  {\em JHEP} {\bf 10} (2014) 178,
  [\href{http://arxiv.org/abs/1407.6429}{{\tt arXiv:1407.6429}}].

\bibitem{Larsen:1995ss}
F.~Larsen and F.~Wilczek, {\it {Internal Structure of Black Holes}},  {\em
  Phys. Lett.} {\bf B375} (1996) 37--42,
  [\href{http://arxiv.org/abs/hep-th/9511064}{{\tt hep-th/9511064}}].

\bibitem{Solodukhin:1995ak}
S.~N. Solodukhin, {\it {One Loop Renormalization of Black Hole Entropy Due to
  Nonminimally Coupled Matter}},  {\em Phys. Rev.} {\bf D52} (1995) 7046--7052,
  [\href{http://arxiv.org/abs/hep-th/9504022}{{\tt hep-th/9504022}}].

\bibitem{Solodukhin:1996jt}
S.~N. Solodukhin, {\it {Nonminimal Coupling and Quantum Entropy of Black
  Hole}},  {\em Phys. Rev.} {\bf D56} (1997) 4968--4974,
  [\href{http://arxiv.org/abs/hep-th/9612061}{{\tt hep-th/9612061}}].

\bibitem{Hotta:1996cq}
M.~Hotta, T.~Kato, and K.~Nagata, {\it {A Comment on Geometric Entropy and
  Conical Space}},  {\em Class. Quant. Grav.} {\bf 14} (1997) 1917--1925,
  [\href{http://arxiv.org/abs/gr-qc/9611058}{{\tt gr-qc/9611058}}].

\bibitem{Akers:2015bgh}
C.~Akers, O.~Ben-Ami, V.~Rosenhaus, M.~Smolkin, and S.~Yankielowicz, {\it
  {Entanglement and RG in the $O(N)$ Vector Model}},  {\em JHEP} {\bf 03}
  (2016) 002, [\href{http://arxiv.org/abs/1512.00791}{{\tt arXiv:1512.00791}}].

\bibitem{Banerjee:2015tia}
S.~Banerjee, Y.~Nakaguchi, and T.~Nishioka, {\it {Renormalized Entanglement
  Entropy on Cylinder}},  {\em JHEP} {\bf 03} (2016) 048,
  [\href{http://arxiv.org/abs/1508.00979}{{\tt arXiv:1508.00979}}].

\bibitem{Kay:1990cr}
B.~S. Kay and U.~M. Studer, {\it {Boundary Conditions for Quantum Mechanics on
  Cones and Fields Around Cosmic Strings}},  {\em Commun. Math. Phys.} {\bf
  139} (1991) 103--140.

\bibitem{Kabat:1995eq}
D.~N. Kabat, {\it {Black Hole Entropy and Entropy of Entanglement}},  {\em
  Nucl.Phys.} {\bf B453} (1995) 281--302,
  [\href{http://arxiv.org/abs/hep-th/9503016}{{\tt hep-th/9503016}}].

\bibitem{Blanco:2013joa}
D.~D. Blanco, H.~Casini, L.-Y. Hung, and R.~C. Myers, {\it {Relative Entropy
  and Holography}},  {\em JHEP} {\bf 1308} (2013) 060,
  [\href{http://arxiv.org/abs/1305.3182}{{\tt arXiv:1305.3182}}].

\bibitem{Rosenhaus:2014woa}
V.~Rosenhaus and M.~Smolkin, {\it {Entanglement Entropy: a Perturbative
  Calculation}},  {\em JHEP} {\bf 12} (2014) 179,
  [\href{http://arxiv.org/abs/1403.3733}{{\tt arXiv:1403.3733}}].

\bibitem{Bhattacharya:2012mi}
J.~Bhattacharya, M.~Nozaki, T.~Takayanagi, and T.~Ugajin, {\it {Thermodynamical
  Property of Entanglement Entropy for Excited States}},  {\em Phys.Rev.Lett.}
  {\bf 110} (2013), no.~9 091602, [\href{http://arxiv.org/abs/1212.1164}{{\tt
  arXiv:1212.1164}}].

\bibitem{Nozaki:2013wia}
M.~Nozaki, T.~Numasawa, and T.~Takayanagi, {\it {Holographic Local Quenches and
  Entanglement Density}},  {\em JHEP} {\bf 1305} (2013) 080,
  [\href{http://arxiv.org/abs/1302.5703}{{\tt arXiv:1302.5703}}].

\bibitem{Nozaki:2013vta}
M.~Nozaki, T.~Numasawa, A.~Prudenziati, and T.~Takayanagi, {\it {Dynamics of
  Entanglement Entropy from Einstein Equation}},  {\em Phys.Rev.} {\bf D88}
  (2013), no.~2 026012, [\href{http://arxiv.org/abs/1304.7100}{{\tt
  arXiv:1304.7100}}].

\bibitem{Klebanov:2011uf}
I.~R. Klebanov, S.~S. Pufu, S.~Sachdev, and B.~R. Safdi, {\it {R\'enyi
  Entropies for Free Field Theories}},  {\em JHEP} {\bf 1204} (2012) 074,
  [\href{http://arxiv.org/abs/1111.6290}{{\tt arXiv:1111.6290}}].

\bibitem{Camporesi:1990wm}
R.~Camporesi, {\it {Harmonic Analysis and Propagators on Homogeneous Spaces}},
  {\em Phys.Rept.} {\bf 196} (1990) 1--134.

\bibitem{Bytsenko:1994bc}
A.~A. Bytsenko, G.~Cognola, L.~Vanzo, and S.~Zerbini, {\it {Quantum Fields and
  Extended Objects in Space-Times with Constant Curvature Spatial Section}},
  {\em Phys. Rept.} {\bf 266} (1996) 1--126,
  [\href{http://arxiv.org/abs/hep-th/9505061}{{\tt hep-th/9505061}}].

\bibitem{Smirnov:2006ry}
V.~A. Smirnov, {\em {Feynman Integral Calculus}}.
\newblock 2006.

\bibitem{Hertzberg:2010uv}
M.~P. Hertzberg and F.~Wilczek, {\it {Some Calculable Contributions to
  Entanglement Entropy}},  {\em Phys. Rev. Lett.} {\bf 106} (2011) 050404,
  [\href{http://arxiv.org/abs/1007.0993}{{\tt arXiv:1007.0993}}].

\bibitem{Cardy:2013nua}
J.~Cardy, {\it {Some Results on the Mutual Information of Disjoint Regions in
  Higher Dimensions}},  {\em J.Phys.} {\bf A46} (2013) 285402,
  [\href{http://arxiv.org/abs/1304.7985}{{\tt arXiv:1304.7985}}].

\bibitem{Herzog:2014tfa}
C.~P. Herzog and J.~Nian, {\it {Thermal Corrections to R\'enyi Entropies for
  Conformal Field Theories}},  {\em JHEP} {\bf 06} (2015) 009,
  [\href{http://arxiv.org/abs/1411.6505}{{\tt arXiv:1411.6505}}].

\bibitem{Grover:2011fa}
T.~Grover, A.~M. Turner, and A.~Vishwanath, {\it {Entanglement Entropy of
  Gapped Phases and Topological Order in Three Dimensions}},  {\em Phys.Rev.}
  {\bf B84} (2011) 195120, [\href{http://arxiv.org/abs/1108.4038}{{\tt
  arXiv:1108.4038}}].

\bibitem{Klebanov:2012yf}
I.~R. Klebanov, T.~Nishioka, S.~S. Pufu, and B.~R. Safdi, {\it {On Shape
  Dependence and RG Flow of Entanglement Entropy}},  {\em JHEP} {\bf 1207}
  (2012) 001, [\href{http://arxiv.org/abs/1204.4160}{{\tt arXiv:1204.4160}}].

\bibitem{Huerta:2011qi}
M.~Huerta, {\it {Numerical Determination of the Entanglement Entropy for Free
  Fields in the Cylinder}},  {\em Phys.Lett.} {\bf B710} (2012) 691--696,
  [\href{http://arxiv.org/abs/1112.1277}{{\tt arXiv:1112.1277}}].

\bibitem{Fursaev:2013fta}
D.~V. Fursaev, A.~Patrushev, and S.~N. Solodukhin, {\it {Distributional
  Geometry of Squashed Cones}},  {\em Phys.Rev.} {\bf D88} (2013), no.~4
  044054, [\href{http://arxiv.org/abs/1306.4000}{{\tt arXiv:1306.4000}}].

\bibitem{Liu:2012eea}
H.~Liu and M.~Mezei, {\it {A Refinement of Entanglement Entropy and the Number
  of Degrees of Freedom}},  {\em JHEP} {\bf 04} (2013) 162,
  [\href{http://arxiv.org/abs/1202.2070}{{\tt arXiv:1202.2070}}].

\bibitem{Lewkowycz:2012qr}
A.~Lewkowycz, R.~C. Myers, and M.~Smolkin, {\it {Observations on Entanglement
  Entropy in Massive QFT's}},  {\em JHEP} {\bf 1304} (2013) 017,
  [\href{http://arxiv.org/abs/1210.6858}{{\tt arXiv:1210.6858}}].

\bibitem{Adler:1980pg}
S.~L. Adler, {\it {A Formula for the Induced Gravitational Constant}},  {\em
  Phys. Lett.} {\bf B95} (1980) 241.

\bibitem{Zee:1980sj}
A.~Zee, {\it {Spontaneously Generated Gravity}},  {\em Phys. Rev.} {\bf D23}
  (1981) 858.

\bibitem{Adler:1982ri}
S.~L. Adler, {\it {Einstein Gravity as a Symmetry Breaking Effect in Quantum
  Field Theory}},  {\em Rev. Mod. Phys.} {\bf 54} (1982) 729. [Erratum: Rev.
  Mod. Phys.55,837(1983)].

\bibitem{Rosenhaus:2014nha}
V.~Rosenhaus and M.~Smolkin, {\it {Entanglement Entropy Flow and the Ward
  Identity}},  {\em Phys. Rev. Lett.} {\bf 113} (2014), no.~26 261602,
  [\href{http://arxiv.org/abs/1406.2716}{{\tt arXiv:1406.2716}}].

\bibitem{Rosenhaus:2014ula}
V.~Rosenhaus and M.~Smolkin, {\it {Entanglement Entropy, Planar Surfaces, and
  Spectral Functions}},  {\em JHEP} {\bf 1409} (2014) 119,
  [\href{http://arxiv.org/abs/1407.2891}{{\tt arXiv:1407.2891}}].

\bibitem{Muratani:1983hh}
H.~Muratani and S.~Wada, {\it {The Divergent Parts of Quantum Fluctuation in
  the Curved Space from the Adler-Zee Formulae}},  {\em Phys. Rev.} {\bf D29}
  (1984) 637.

\bibitem{Cappelli:1990yc}
A.~Cappelli, D.~Friedan, and J.~I. Latorre, {\it {C Theorem and Spectral
  Representation}},  {\em Nucl.Phys.} {\bf B352} (1991) 616--670.

\bibitem{Ben-Ami:2015zsa}
O.~Ben-Ami, D.~Carmi, and M.~Smolkin, {\it {Renormalization Group Flow of
  Entanglement Entropy on Spheres}},  {\em JHEP} {\bf 08} (2015) 048,
  [\href{http://arxiv.org/abs/1504.00913}{{\tt arXiv:1504.00913}}].

\bibitem{Smolkin:2014hba}
M.~Smolkin and S.~N. Solodukhin, {\it {Correlation Functions on Conical
  Defects}},  {\em Phys. Rev.} {\bf D91} (2015), no.~4 044008,
  [\href{http://arxiv.org/abs/1406.2512}{{\tt arXiv:1406.2512}}].

\bibitem{Hung:2011nu}
L.-Y. Hung, R.~C. Myers, M.~Smolkin, and A.~Yale, {\it {Holographic
  Calculations of R\'enyi Entropy}},  {\em JHEP} {\bf 1112} (2011) 047,
  [\href{http://arxiv.org/abs/1110.1084}{{\tt arXiv:1110.1084}}].

\bibitem{Perlmutter:2013gua}
E.~Perlmutter, {\it {A Universal Feature of CFT R\'enyi Entropy}},  {\em JHEP}
  {\bf 03} (2014) 117, [\href{http://arxiv.org/abs/1308.1083}{{\tt
  arXiv:1308.1083}}].

\bibitem{Bueno:2015qya}
P.~Bueno, R.~C. Myers, and W.~Witczak-Krempa, {\it {Universal Corner
  Entanglement from Twist Operators}},  {\em JHEP} {\bf 09} (2015) 091,
  [\href{http://arxiv.org/abs/1507.06997}{{\tt arXiv:1507.06997}}].

\bibitem{Osborn:1993cr}
H.~Osborn and A.~Petkou, {\it {Implications of Conformal Invariance in Field
  Theories for General Dimensions}},  {\em Annals Phys.} {\bf 231} (1994)
  311--362, [\href{http://arxiv.org/abs/hep-th/9307010}{{\tt hep-th/9307010}}].

\bibitem{Zamolodchikov:1986gt}
A.~B. Zamolodchikov, {\it {Irreversibility of the Flux of the Renormalization
  Group in a 2D Field Theory}},  {\em JETP Lett.} {\bf 43} (1986) 730--732.
  [Pisma Zh. Eksp. Teor. Fiz.43,565(1986)].

\bibitem{Klebanov:2012va}
I.~R. Klebanov, T.~Nishioka, S.~S. Pufu, and B.~R. Safdi, {\it {Is Renormalized
  Entanglement Entropy Stationary at RG Fixed Points?}},  {\em JHEP} {\bf 10}
  (2012) 058, [\href{http://arxiv.org/abs/1207.3360}{{\tt arXiv:1207.3360}}].

\bibitem{Nishioka:2014kpa}
T.~Nishioka, {\it {Relevant Perturbation of Entanglement Entropy and
  Stationarity}},  {\em Phys. Rev.} {\bf D90} (2014), no.~4 045006,
  [\href{http://arxiv.org/abs/1405.3650}{{\tt arXiv:1405.3650}}].

\bibitem{Donnelly:2012st}
W.~Donnelly and A.~C. Wall, {\it {Do Gauge Fields Really Contribute Negatively
  to Black Hole Entropy?}},  {\em Phys. Rev.} {\bf D86} (2012) 064042,
  [\href{http://arxiv.org/abs/1206.5831}{{\tt arXiv:1206.5831}}].

\bibitem{Donnelly:2014fua}
W.~Donnelly and A.~C. Wall, {\it {Entanglement Entropy of Electromagnetic Edge
  Modes}},  {\em Phys. Rev. Lett.} {\bf 114} (2015), no.~11 111603,
  [\href{http://arxiv.org/abs/1412.1895}{{\tt arXiv:1412.1895}}].

\bibitem{Donnelly:2015hxa}
W.~Donnelly and A.~C. Wall, {\it {Geometric entropy and edge modes of the
  electromagnetic field}},  {\em Phys. Rev.} {\bf D94} (2016), no.~10 104053,
  [\href{http://arxiv.org/abs/1506.05792}{{\tt arXiv:1506.05792}}].

\bibitem{Huang:2014pfa}
K.-W. Huang, {\it {Central Charge and Entangled Gauge Fields}},  {\em Phys.
  Rev.} {\bf D92} (2015), no.~2 025010,
  [\href{http://arxiv.org/abs/1412.2730}{{\tt arXiv:1412.2730}}].

\bibitem{Soni:2016ogt}
R.~M. Soni and S.~P. Trivedi, {\it {Entanglement Entropy in (3+1)-D Free $U(1)$
  Gauge Theory}},  \href{http://arxiv.org/abs/1608.00353}{{\tt
  arXiv:1608.00353}}.

\end{thebibliography}\endgroup

\end{document}